\documentclass[titlepage]{article}

\usepackage[utf8]{inputenc}
\usepackage[font=small,labelfont=bf]{caption}%
\usepackage {graphicx}
\usepackage{indentfirst}
\usepackage[usenames]{color}
\usepackage[unicode,colorlinks]{hyperref}
\usepackage{natbib}

\usepackage{caption}
\usepackage{authblk}

\def\aapr{The Astronomy and Astrophysics Review}
\def\apj{Astrophys.~J}

\def\aap{Astron.~Astrophys}

\def\nat{Nature}
\def\apjl{Astrophys.~J.~Lett}
\def\apjs{Astrophys.~J.~Suppl.~Ser}
\def\aj{Astron.~J}
\def\mnras{Mon.~Not.~R.~Astron.~Soc}

\begin{document}
\title{The Spatial---Kinematic Structure of the Region of Massive Star
	Formation S255N on Various Scales
}
\author[1]{P. M. Zemlyanukha}
\author[1,2]{I. I. Zinchenko}
\author[3]{S. V. Salii}
\author[1,2]{O. L. Ryabukhina}
\author[4]{S.-Y. Liu}
\affil[1]{Institute of Applied Physics, Russian Academy of Sciences, Nizhnyi Novgorod, Russia}
\affil[2]{Lobachevsky State University, Nizhnyi Novgorod, Russia}
\affil[3]{B.N. Yeltsin Ural Federal University, Ekaterinburg, Russia}
\affil[4]{Institute of Astronomy and Astrophysics, Academia Sinica, P.O. Box 23-141, Taipei 106, Taiwan}
\maketitle

\def \co {C$^{18}$O (2---1)}
\def \nh {N$_2$H$^+$ (3---2)}
\def \nhh {NH$_3$ (1,1)}
\def \so {SO (6$_{5}$---5$_{4}$)}

ASTRONOMY REPORTS Vol. 62 No. 5 2018

\section*{Abstract}
The results of a detailed analysis of SMA, VLA, and IRAM observations of the region of massive star formation S255N in CO(2---1), \nh, \nhh, \co  and some other lines is presented. Combining interferometer and single-dish data has enabled a more detailed investigation
of the gas kinematics in the moleclar core on various spatial scales. There are no signs of rotation or
isotropic compression on the scale of the region as whole. The largest fragments of gas ($\approx$0.3 pc) are located
near the boundary of the regions of ionized hydrogen S255 and S257. Some smaller-scale fragments are
associated with protostellar clumps. The kinetic temperatures of these fragments lie in the range 10---
80 K. A circumstellar torus with inner radius R$_{in}$ $\approx$ 8000 AU and outer radius R$_{out}$  12 000 AU has been
detected around the clump SMA1. The rotation profile indicates the existence of a central object with mass
$\approx$ 8.5/ sin 2 (i) M$_\odot$ . SMA1 is resolved into two clumps, SMA1---NE and SMA1---SE, whose temperatures
are $\approx$150 K and $\approx$25 K, respectively. To all appearances, the torus is involved in the accretion of surrounding
gas onto the two protostellar clumps.

\section{Introduction}
Studies of molecular clouds have attracted considerable attention. Much effort has gone into both
simulations and observations of these objects (see, e.g., \citep{rotcluster,obs}). However, there remain relatively few well studied regions of high-mass star formation. They are encountered more rarely than the regions of low-mass star formation, and are located at large distances from the Sun, complicating observations of them. In relation to detailed studies of processes occurring in their cores, the resolution of single dishes is insufficient, while interferometers have limited sensitivity to extended structures.

\begin{figure}[h]
	\includegraphics[width=0.9\textwidth]{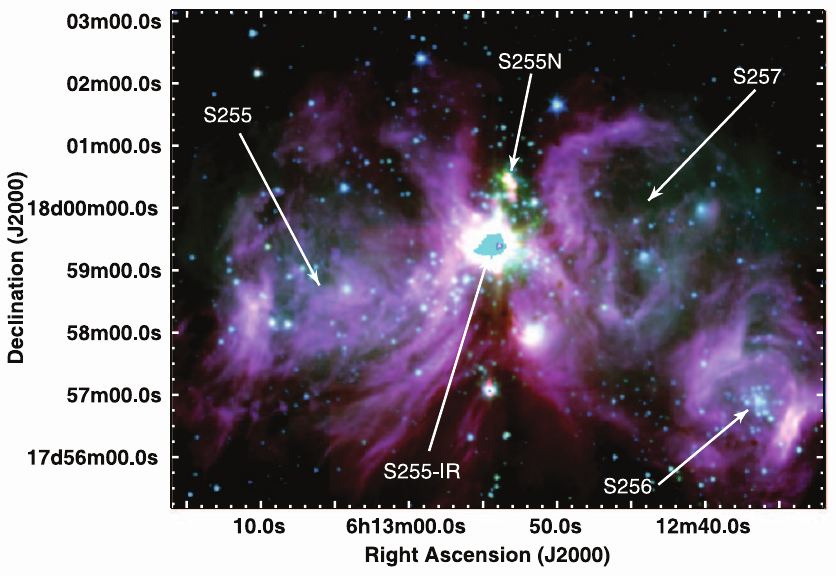}
	\caption{Spitzer IRAC three-color image of the region S255N and its surroundings, composed of images at 8.0 $\mu$m (red), 4.5 $\mu$m (green), and 3.6 $\mu$m (blue). S255N is located in a complex of regions of current and past star formation \citep{cyg2007}. The color figures here and below are accessible in the electronic version of the journal.}
	\label{fig:cyg}
\end{figure}

The object considered in the current study is part of a molecular cloud \citep{oldie,lev} located between the two zones of ionized hydrogen S255 and S257 (Fig.~\ref{fig:cyg}). It is supposed that the entire S254---S258 complex was formed by successive star formation\citep{bieging}: expansion ofthe HII zones led to the compression of gas in the molecular cloud. This cloud contains three cores: S255IR, S255S, and S255N. The first is located in a later stage of evolution \citep{wang,ojha}: its radiation is appreciably brighter in the IR, and water-vapor and Class II methanol maser emission is detected in this
region \citep{zinhr,kurtz}. The core S255S is the youngest of the three, as is testified to by its lower brightness in the IR and in the 1.3~mm lines, as well as the compactness of the outflow in the core \citep{wang}. A smaller number of 1.3~mm molecular lines is detected [6], and there are no signs of massive-star formation; we accordingly did not consider it in the current study.

The molecular cloud with which S255N (Sh2-255 FIR1, G192.60-MM1) is associated has been observed on several single-dish radio telescopes (OSO 20 m, IRAM 30 m, NRAO 12 m) \citep{zin2009}; the core of S255N has been observed with the SMA and VLA interferometers \citep{zinhr}. We present here combined data for the first time. The mass of the S255N core indicated by single-dish observations is $\sim$300 M$_\odot$, with  n$_{H_{2}}$~$\sim$2$\times$10$^5$ cm$^{-3}$, T$_k$~$\sim$40K, $\Delta$V~$\sim$2km/s. The luminosity of the core is of order 10$^5$ L$_\odot$ \citep{minier}. An ultracompact HII region is observed in this region\citep{kurtz94}, as well as H$_2$O and Class I CH$_3$OH masers. These characteristics of the core provide evidence for the formation of massive stars in this region. A lack of coincidence in the positions of the emission peaks in the CO(2---1), HCN(1---0), HNC(1---0), HCO$^+$(1---0), \co, and C$^{34}$S(5---4) molecular lines has also been demonstrated \citep{zin2009}. The presence of hot, massive protostellar clumps in the core was shown in \citep{zinhr,cyg2007}, with the velocity of half of the cores are different from the velocity of the ambient gas. The mass of the brightest of these clumps is $\approx$16 M$_\odot$ \citep{zinhr} for a distance of 1.78 kpc\citep{burns}. The presence of two spectral components in the (1, 1) ammonia lines in directions not coincident with a hot clump was detected in\citep{zinhr}. Several high-velocity outflows are present in the core, which also indicates rich gas kinematics. We attempted in the current study to estimate the motion of gas on scales fom the size of the core to the sizes of clumps, based on the collected data. We also used non-standard data-analysis methods that are robust to a noise to analyze emission that is weak against the background of the hot-clump emission. We used original methods to detect and identify kinematic fragments of the core, which were traced in position–velocity diagrams in a number of molecular lines.

\section{Observational data and analysis} 
We considered data obtained on the SMA\citep{wang,zinhr} and VLA\citep{zinhr} interferometric arrays and the 30-m IRAM telescope \citep{wang,zinhr}. Table~\ref{t:mol} presents a list of the observed molecules, the frequencies of their transitions, and the instruments used. All the data were reduced anew, and self-calibration was appied to the data of \citep{zinhr}. The CO(2---1), \co, CH$_3$CN(12---11), and SO(6$_{5}$---5$_{4}$) lines were observed twice in the compact and once in the extended configuration of the SMA. These data were inverted into a single image at each line. The width of the synthesized antenna beam was  $\approx$ 1.14$''$ at 217 GHz and $\approx$ 3.4$''$ for the remaining observations (the compact configuration of the SMA). The width of the VLA beam was $\approx$ 2.55$''$ (23.7 GHz), and the width of the IRAM beam was 12$''$ (217 GHz). The coordinates of the phase center for the SMA observations were $\alpha$(2000)=06$^h$12$^m$53.800$^s$, $\delta$(2000)=17$^\circ$59$'$22.097$''$.

\subsection{Data reduction}
Since one of the main aims of our study was to investigate the kinematics of the protostellar core on various spatial scales, we considered objects detected earlier in the 255N core with various extents from 2.5$\times$10$^{-3}$ pc to 0.48 pc [11]. The maximum angular sizes to which the instruments used at the various frequencies are sensitive are 66$''$ for the VLA and 25$''$ for the SMA, which corresponds to linear sizes of 0.52 and 0.2 pc at a distance of 1.78 kpc \citep{minier}. These limits hinder the interpretation of observations over a wide range of spatial scales. An important role is also played by the loss of flux by the interferometer, which we compensated for by combining the interferometric and single-dish data. As an example, the profile of the CO(2---1) line indicated by the SMA observations differs strongly from the profile observed on the IRAM 30-m telescope in the same direction. Therefore, we consolidated the SMA data with the observations on
the 30-m IRAM antenna. This was carried out for the \nh, CO(2---1), and SiO(5---4) lines, as well as the 1.3-mm continuum observations. The spectral-line data were combined in the visibility domain based on the relative integrated-flux calibration
in the region of overlapping spatial frequencies; the reconstructed continuum images were consolidated via a Fourier transform using standard procedures of the MIRIAD package \citep{miriad}. The “wings” of the CO(2---1) line at 3--7 km/s and 11--15 km/s from the single-dish data were used for the relative flux calibration of the synthesized visibilities at the frequency of this line.

CO(2---1), C$^{18}$O(2---1), SO(6$_{5}$---5$_{4}$), DCO$^+$(4---3), DCN(3---2) data were self-calibrated using the observations in the CH$_3$OH ($4_{2}$-$3_{1}$) line \citep{zinhr}. The self-calibration based on this line was
fairly effective, since a single maser source was observed in the telescope field \citep{selfcal}.

\subsection{Identification of kinematic fragments of the core}
The presence of at last two spectral components with different Doppler velocities in the 255N region
was noted earlier \citep{zinhr}, however their spatial distribution was not studied. In our current study, we carried out a careful analysis of the data considered, with the aim of conducting a detailed study of the spatial---kinematic structure of the core. 

\begin{figure}[h]
	\includegraphics[width=0.9\textwidth]{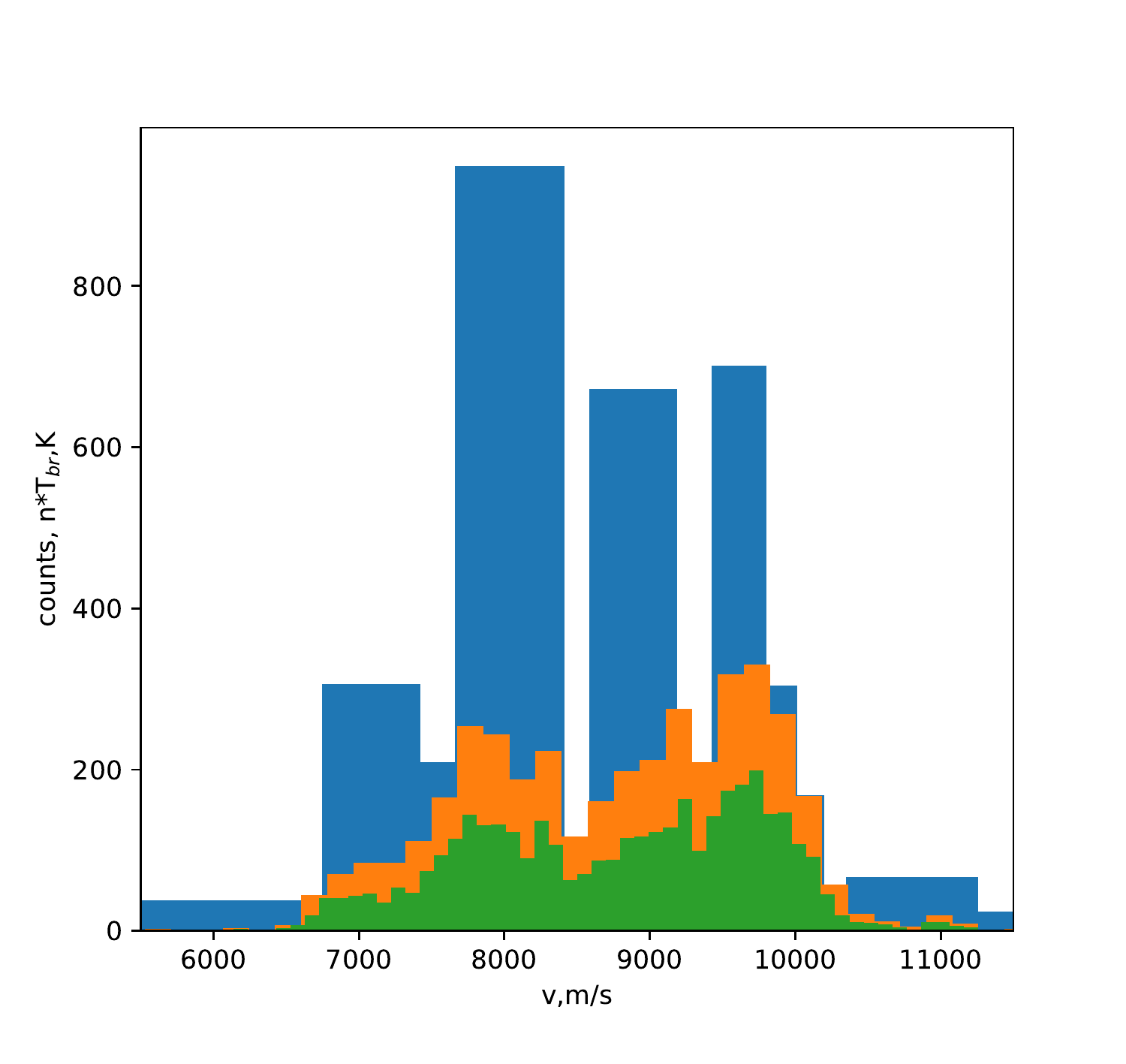}
	\caption{Example of the velocity distribution of the \nhh intensities derived using various algorithms for choosing the bin width. The blue columns show the results for the Bayesian block method, the orange bins the results of applying the Knuth rule, and the green columns results for a bin width of 0.2 km/s.}
	\label{fig:hist}
\end{figure}

Application of the moment method for the identification of kinematic fragments does not always provide the best solution, especially for lines of molecules with hyperfine structure, such as NH$_3$ and N$^2$H$^+$,
since such estimates can be shifted due to asymmetry of the line caused by its hyperfine structure. In addition, in the case of observations of two or more spectral components, the first moment yields an estimate of their mean velocity. Determining the spatial boundaries of components precisely is problematic, since they can interact, overlap, and be represented differently in different molecular lines, depending on whether or not the conditions in the components are the same. Therefore, we estimated the minimum number of boundaries from the number of peaks in a histogram of the line intensities as a function of their velocities (Fig.~\ref{fig:hist}). Thus, instead of a spatial distribution, we obtained the distribution of the line intensities in various velocity ranges. This characterizes the distribution of the density, temperature, or some other physical parameter, depending on what quantity is traced by the studied line. We chose three molecular lines for this analysis: CO(2---1), \nh, and \nhh. We expected to observe signs of fragments in the outer, rarified layers of the core in the CO line. The emission in NH$_3$ and N$^2$H$^+$ is excited in quiescent and dense regions of the core. It is possible to obtain more accurate velocity estimates using these lines, since their transitions display hyperfine structure. Their spectra are fitted fairly well with the local thermodynamic equilibrium (LTE) model presented below. The shape of the histogram (Fig.~\ref{fig:hist}) provides information about the internal structure of the core -- whether it is homogeneous or is appreciably perturbed, and also whether the gas can be described as forming isolated or overlapping velocity components (that is, as kinematic fragments of the core), or whether the gas is involved in some process with a single spectral distribution.

We estimated the positions of the line centers by fitting model line profiles for the transition to the observed spectra, taking into account hyperfine structure in an LTE approximation with a Gaussian optical-depth profile:

\begin{equation}
T(\nu)=\sum_{n_g=1}^{m} (J(T_{ex}^{n_g}) - J(T_{bg}))(1-e^{-\tau_{n_g}(v)}),
\end{equation}
\begin{equation}
\tau_{n_g}(v) = \sum_{i=1}^{n_{hfs}} 16\pi^3\nu_{ref}^4\mu^2 \frac{S g_i N_{u}^{n_g} } {3 c^3 J(T_{ex}^{n_g}) } \phi_{n_gi}(v),
\end{equation}
\begin{equation}
\phi_{n_gi}(v) = \frac{1}{\sqrt{2 \pi} \sigma_{n_g} } e^{\frac{-(v-v_i-v_{n_g})^2}{2 \sigma_{n_g}^2}}, 
\end{equation}
\noindent where $T_{ex}^{n_g}$ is the excitation temperature of spectral component $n_g$, $N_{u}^{n_g}$ the column density of the molecules in the upper level for the transition considered,  $\sigma_{n_g}$ the line width, $v_{n_g}$ the position of the line-component center (the subscript $n_g$ denotes the desired parameters of the model used in the fitting), $J(T)=\frac{h\nu/k}{exp(h\nu/kT)-1}$, $T_{bg} = 2.73 K$  is the cosmic microwave background temperature, $\nu_{ref}$ the transition frequency, $\mu$ the dipole moment, $S$ the line strength, $g_i$ the statistical weight of transition i of the hyperfine structure, $n_{hfs}$ the number of components of the hyperfine structure (18 for NH$_3$(1,1), 24 for NH$_3$(2,2), 38 for N$_2$H$^+$(3---2)) $v_i$ the Doppler velocity of the transition $i$ in the line, and $m$ the number of spectral components in the line.

The values of $S$,$g_i$ and$v_i$  were taken from the CDMS database \citep{cdms}. The approximation was applicable for the NH$_3$ (1,1) and (2,2) and N$_2$H$^+$(3---2) lines. At a number of positions on the map, two peaks were observed in the spectrum; accordingly, we allowed for the possibility of fitting lines containing $m$ independent spectral components. In our data sets, $m$ did not exceed two. The choice between one or two components was based on a comparison of the rms deviations in the spectrum after subtracting the model. If the ratio of the deviations for $m$=2 and $m$=1 was less than 1.4, we used the two-component model.

Thus, after fitting a line at each point in the map, we had estimates of the Doppler shifts of one or two spectral components. However, the number of kinematic components over the entire map can be appreciably greater than two. The number of such components can be estimated from a histogram of the velocity distribution of the line intensities. In the simplest case, the integrated intensities of the spectral components whose Doppler velocities lay within a given bin of the histogram were added. There also exist a large number of methods enabling choice of the bin width based on the data set considered, two
of which we used in our analysis: the rule of Knuth \citep{knuth} and the Bayesian block method \citep{bb}, as realized
in the astropy library \citep{astropy}. Figure~\ref{fig:hist} presents examples of applying these various methods. 
	
\begin{figure}[t]
	\includegraphics[width=.9\textwidth]{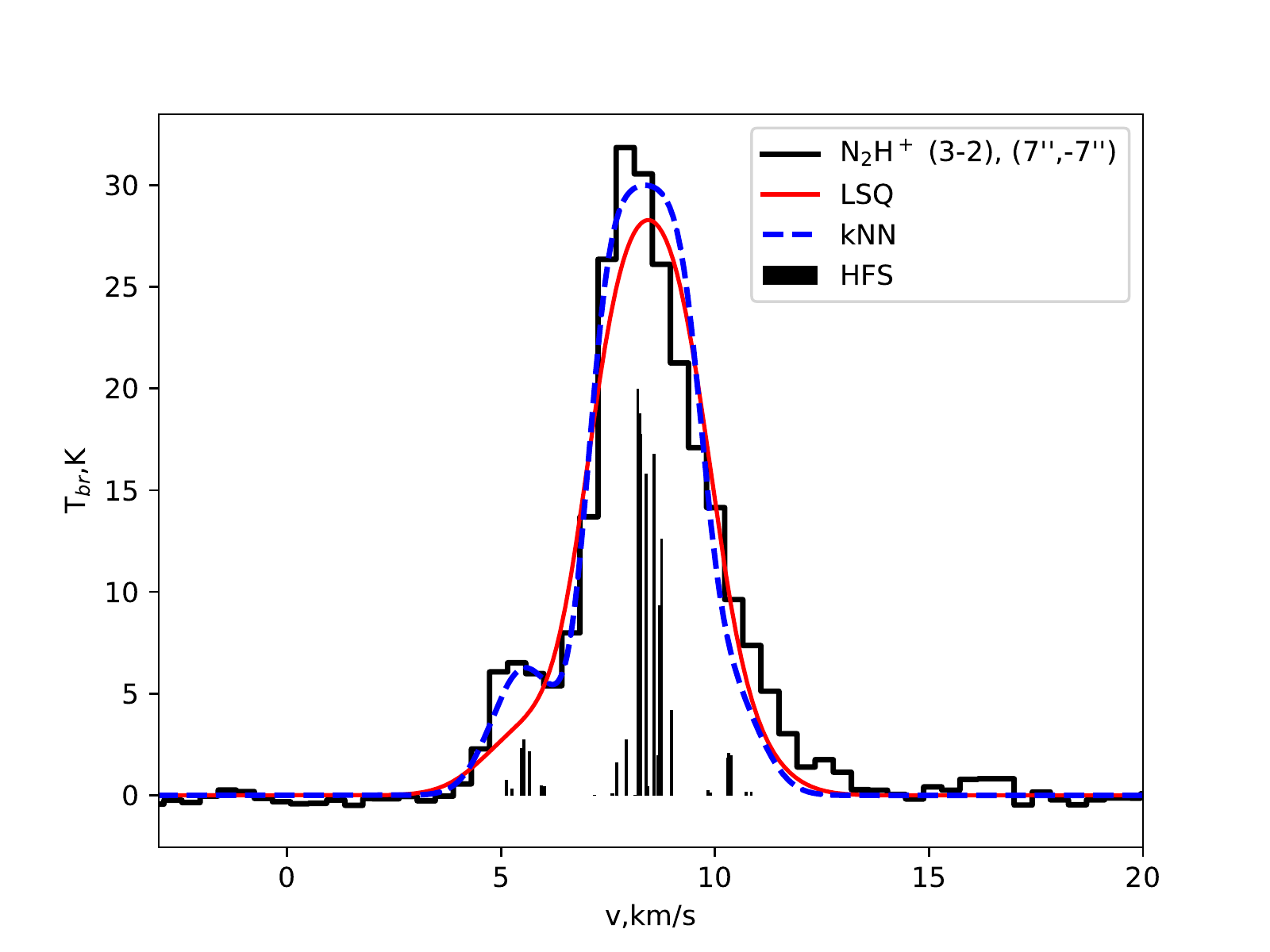}
	\caption{Observed \nh spectrum at the point (-7$''$ , 7$''$) relative to S225N--SMA1 (histogram), together with fits obtained using the least squares method (red solid curve) and the kNN method (blue dashed curve). The vertical lines with lengths proportional to their statistical weights denote hyperfine components of the (3---2) transition.}
	\label{fig:fit}
\end{figure}

Since a large fraction of a map contains lines with intensities at the detection threshold (3---5$\sigma$), and the channel width is of order the line width (5 in a group, 25 in a line, in the case of ammonia), we estimated the velocity using the k nearest neighbors (kNN) method \citep{knn}, since least-squares fitting for these regions either diverged or converged to a local minimum, with the suppression of one component by the other. Based on the fitting at each point in the map, regions were chosen whose $\chi^2$ values for the LTE approximation were less than unity. The spectra at these points were used as a reference set of objects relative to which the remaining regions were considered. The velocities were estimated as

\begin{equation}
v_k=\frac {\sum_{i=1}^n v_i,d(s_k,s_i)^2}{\sum_{i=1}^n d(s_k,s_i)^2}
\end{equation}
where $v_k$ is the velocity in a spectrum of interest, $v_i$ the velocity for the known (reference) spectrum, $s_k$, $s_i$ a spectrum of interest and a spectrum whose estimated velocity is known, $n=5$ the number of nearby spectra used to estimate the velocity, $d(s_1,s_2) = \sqrt{\sum_{j=1}^m (s_{1_j} - s_{2_j})^2}$ the distance in intensity space between the two spectra (analogous to the least-squares residual), $m$ the number of channels, and $s_{i_j}$ the intensity of the spectrum in channel $j$. In this form, the method resembles least squares fitting, but the algorithm for the descent of the parameters to those yielding the minimum residual is replaced by a search of n "similar" spectra from the reference set. Examples of fitting a spectrum using the least-squares method and the first nearest neighbor method are presented in Fig.~\ref{fig:fit}. The $\chi^2$ value for the LTE approximation is 266 (Jy/beam)$^2$ , and for the first nearest neighbor method is 72 (Jy/beam) $^2$. A more detailed analysis of the k nearest neighbors method will be described in future publications.

\subsection{Analysis of methanol observations}

Estimates of the physical parameters (kinetic temperature and density of molecular hydrogen, specific column density (column density of molecules per unit velocity interval), and relative methanol abundance) were carried out using the database of methanol energy-level populations \citep{sali2004}. This database contains methanol energy-level populations calculated at nodes of a parameter grid. The parameter values were varied from those corresponding to a dark molecular cloud to values that may be realized in regions heated by an embedded object or excited by the passage of a shock. The kinetic temperature was varied from 10 to 220 K, the density of molecular hydrogen from  $10^3$ to $10^9$ cm$^{-3}$, the abundance of methanol relative to hydrogen from  $10^{-9}$ to $10^{-6}$, and the specific methanol column density from $10^8$ to $10^{13}$ cm$^{-3}$s. The computations of the methanol level populations in the database were carried out in the LVG approximation. A model for the excitation of the methanol molecule developed for the conditions characteristic for star-forming regions, including regions of massive-star formation, was used in the computations \citep{sobolev}.

\section{Results}
\subsection{Spatial--kinematic structure of the core}
To investigate the kinematics of the gas inside the core, we considered first and foremost the emission in the \nh and \nhh lines, since these trace dense, quiescent gas. In addition, both lines contain hyperfine structure, making it possible to accurately measure Doppler shifts, and thereby the velocity of the gas. We compared the \nh data with the data for the CO(2---1), (6$_{5}$---5$_{4}$), and \co lines in a position–velocity diagram (see Fig. \ref{fig:pvcore}).

\begin{figure}[t]
	\includegraphics[width=.9\textwidth]{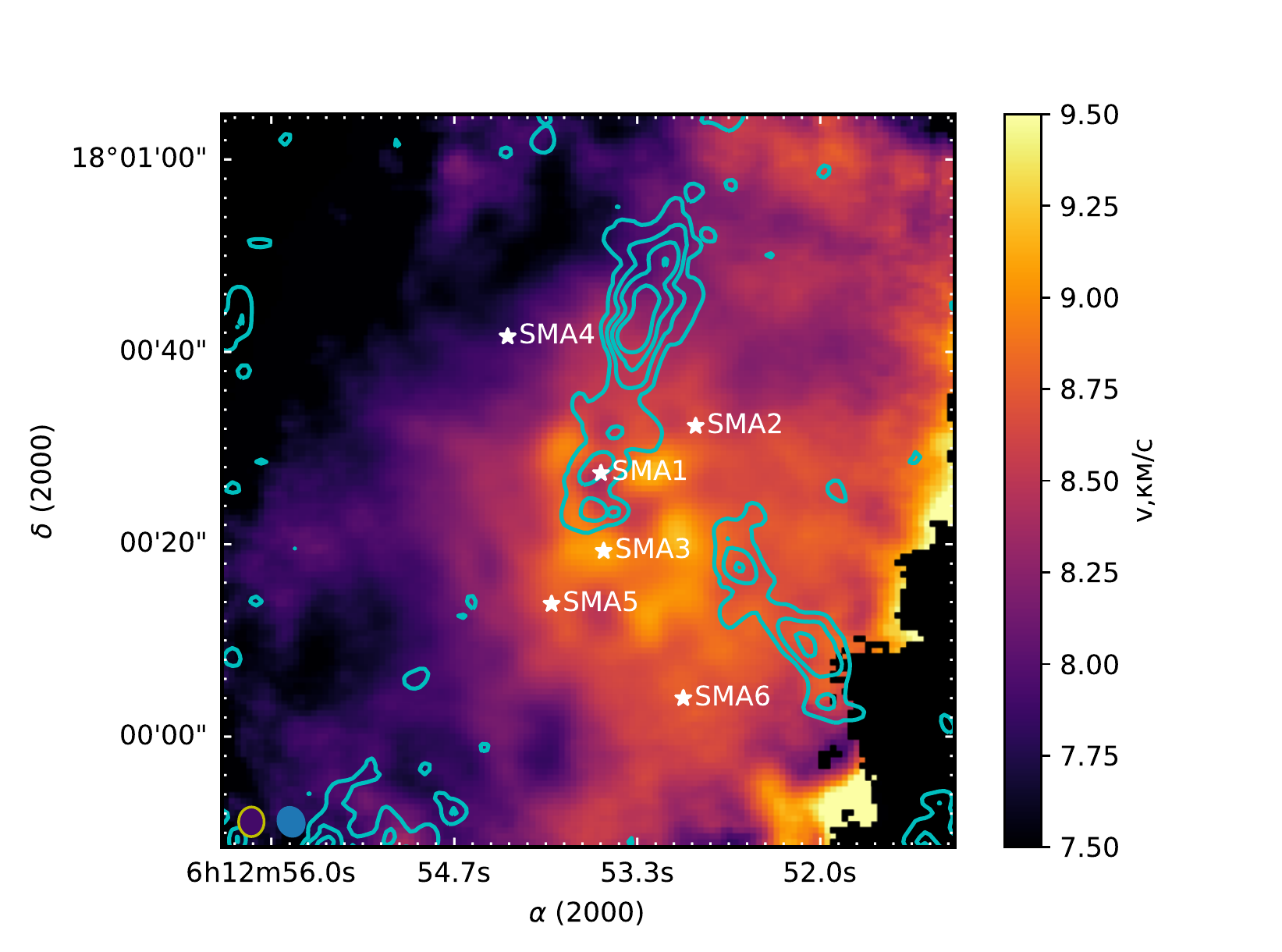}
	\caption{Map of the first moment of the CO(2---1) line (background). Contours of the (1, 1) ammonia intensity at levels of [4, 5.4, 6.8, 8, 9.5] K are superposed. The stars show the positions of clumps according to the data of \citep{zinhr}.}
	\label{fig:mco}
\end{figure}

\begin{figure}[h]
	\includegraphics[width=.9\textwidth]{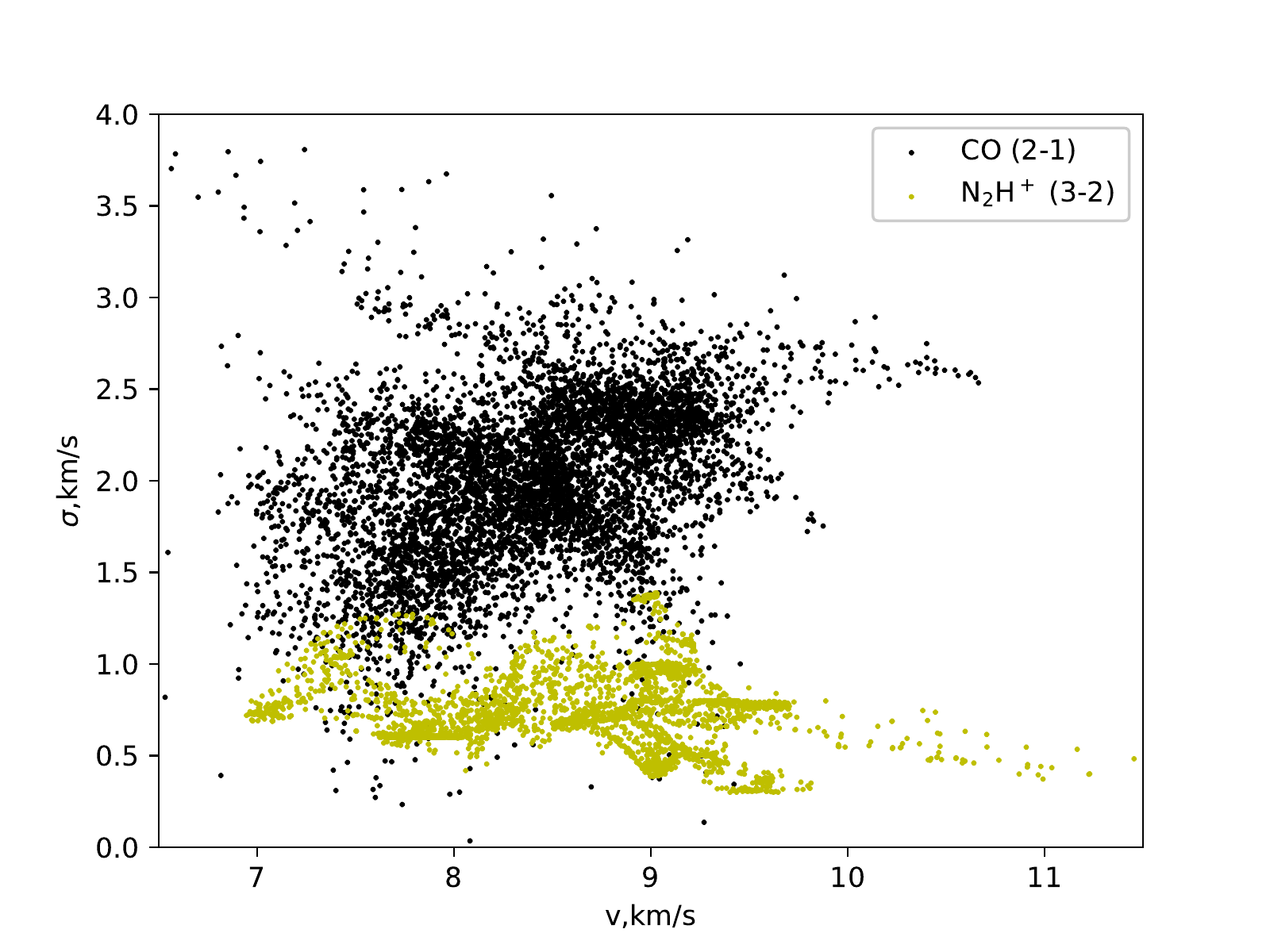}
	\caption{Dependence of the second moment (width) on the first moment (velocity) of the CO (2---1) and \nh lines.}
	\label{fig:cocorr}
\end{figure}

The S255N core is part of a more extended structure \citep{oldie}, which is also observed in our data. The CO(2---1) line emission has no boundary at the north and south of the map, as can be seen in Fig. \ref{fig:mco} (combined data). The map of the first moment of this line on scales exceeding 20$''$ corresponds to the map presented in \citep{lev}. However, a comparison of the first and second moments of the CO line (Fig. \ref{fig:cocorr}) indicates that the line width grows toward the center of the velocity range in a close to linear shape. The direction of the velocity variations is close to the direction toward the ionized regions S255 and S257 from the center of the map, with the mean gas velocity observed at the map
center. There is no clearly distinguished symmetry axis, such as would be characteristic for the presence of rotation, as was shown in \citep{rotcluster}. Such a distribution would not be characteristic for an isotropic collapse. However, the broadening of lines toward the center of the velocity range may indicate mixing of two portions of gas driven by the expansion of envelopes.

\begin{figure}[t]
	\includegraphics[width=.9\textwidth]{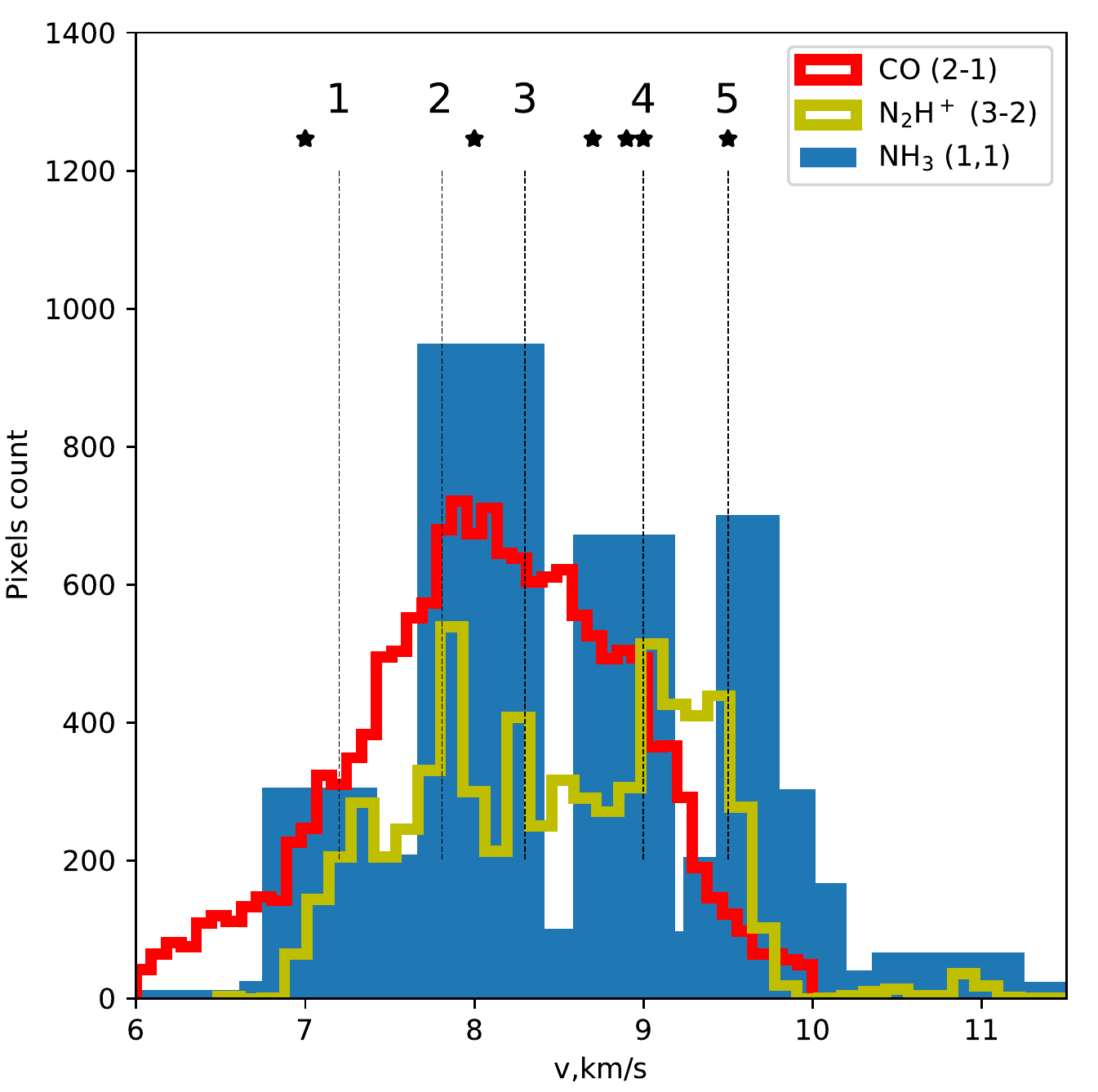}
	\caption{Histograms of the velocity distributions in various lines. The numbers and vertical dotted lines denote bands corresponding to the positions of kinematic fragments of the core. Maps of the first moment were used to construct the CO velocity histogram, and velocity estimates obtained with the kNN method were used for the remaining lines. The stars denote the velocities of clumps according to the data of \citep{zinhr}.}
	\label{fig:nvh}
\end{figure}

As can be seen in Fig. \ref{fig:kn2h}, the kinematics of the core are very complex. The presence of two spectral components of ammonia in gas that is not associated with protostellar clumps was indicated in \citep{zinhr}, as we also found in our data. Shifts in the component velocities from point to point are observed. The overall direction of the velocity gradient lies from the Southwest to the Northeast and East, from $\sim$6.7 km/s to $\sim$11 km/s. However, the velocity does not vary uniformly. There are regions where there is essentially no shift in the line. Figure 7 presents the velocity distribution of the pixels for maps of the CO(2---1), N$_2$H$^+$(3---2) and NH$_3$(1,1) lines. The greatest coincidence of peaks in the distributions for the various lines occurs at a velocity of $\sim$7.8 km/s. The gas with such velocities is localized primarily in the southwestern part of the map. There is also a modest peak at $\sim$7.1 km/s located at the southern edge of the map, which can also be identified in the velocity slice marked 1 in Fig. \ref{fig:pvcore}. According to the data of \citep{lev,zinhr}, the velocity of the S255N core is 8.9 km/s, which corresponds to peak 4 in Fig. \ref{fig:nvh}. However, its position in CO is shifted by $\sim$0.3 km/s relative to its position in N$_2$H$^+$ and NH$_3$ . Modest peaks are also present at 8.3 km/s and 9.6 km/s. We denote each kinematic fragment using numbers from 1 to 5, beginning with the fragment with the lowest velocity, according to the histogram in Fig. \ref{fig:nvh}. 

Thus, the gas in the core is distributed inhomogeneously. There exist some regions within which there are no significant velocity shifts, which we will refer to as kinematic fragments of the core. Figure \ref{fig:pvcore} presents a frequency slice through the data cube in the CO(2---1), N$_2$H$^+$(3---2), SO(6$_{5}$---5$_{4}$) and C$^{18}$O(2---1) lines along a path passing along the periphery of the core and crossing through these fragments. This path is shown by the red line in Fig.\ref{fig:nh3_pv}. The CO spectrum slice shape overall repeats the spatial–frequency structure of the N$_2$H$^+$ and NH$_3$ lines, however, appreciably more compact structures are traced in the SO and C$^{18}$O lines. Fragments 1, 3, and 4 are correlated with both of these lines, while fragment 2 is predominantly correlated with C$^{18}$O. CO emission at $\sim$7.9 km/s is observed along the entire path of the slices. We suggest that this is associated with the diffuse gas surrounding the core observed in \citep{oldie}. In contrast to CO, the  N$_2$H$^+$ and NH$_3$ emission at this velocity is localized only in the southeastern part of the core. This region is depicted in Fig. \ref{fig:pvcore} at the beginning and end of the slice path. Two spectral components in both the  N$_2$H$^+$ and NH$_3$ lines are observed a distance 50$''$ from the beginning of the path, as was also noted earlier in \citep{zinhr}. The red component is not obviously represented in the CO line. There is likewise no boundary between the fragments in the CO and N$_2$H$^+$ lines with angular and spectral resolutions of 1.2$''$ and 0.5 km/s, apart from the fifth fragment. It is also noteworthy that, in spite of the different spatial positions of the emission maxima in the N$_2$H$^+$ and NH$_3$ lines, their spatial--frequency distributions are otherwise similar to each other. The large extent of the N$_2$H$^+$ emission can be explained by the addition of the single-dish data to the interferometric data.

The velocities of the clumps SMA1, SMA2, and SMA3 presented in \citep{zinhr} are close to the gas velocities observed in the \nh line, which is not the case for SMA4, SMA5, and SMA6. In addition, a continuous, elongated filament of gas in the \co line can be traced between clumps SMA1--3; the velocity gradient along this filament is reflected in the position–velocity diagram in Fig. \ref{fig:psma123}. It is striking that the DCO$^+$ (4---3) line is shifted relative to the of the \co emission peak observed in the clumps. The peak of the DCO$^+$ emission, which is located at the edge of the map, has coordinates (7$''$, --34$''$) relative to the phase center, and a full width at half maximum of 0.7 km/s with its center at 8.8 km/s and an intensity of $\approx$3 K. The DCO$^+$, CO, and N$_2$H$^+$ spectra are presented in Fig. \ref{fig:dcospecs}. The CO line has red wing, suggesting a high-velocity outflow. An IR source with similar coordinates is presented in \citep{ir,spitzer}, which in all likelihood is identified with an outflow and the DCO peak. The spatial–velocity structure of the gas is more complex in the vicinity of the clump SMA1 than in the rest of the core. 

\subsection{Central region}
The region of intersection of kinematic fragments 2, 3, and 5 in the vicinity of SMA1 is of the most interest. The peak of the 1.3-mm continuum emission also lies in this direction. According to the data presented in \citep{zinhr}, the Doppler velocity of the brightest clump, associated with a peak in the dust emission of S255N---SMA1, is 8 km/s; however, the presence of two spatially unresolved clumps inside SMA1 is proposed in \citep{cyg2007}. The existence of a bipolar outflow in the CO(2---1) and SiO(5---4) lines is also known. Evidence for rotation of the clump SMA1 with a radius of $\sim$1.5$''$, corresponding to 2500 AU at a distance of 1.78 kpc, is presented in \citep{wang}, but the hypothess that two clumps are present was not considered. We observed an appreciably more extended structure toward SMA1 in the ammonia data (Fig. \ref{fig:nh3}) than the structure described in \citep{wang,zinhr}, oriented perpendicular to the outflow, whose velocity varies from 9 to 10.4 km/s, primarily along the structure. Figure 12 presents a velocity map together with a position-velocity diagram for this clump in the \co and DCO$^+$ (4---3) lines with the velocities estimated from the ammonia line indicated.

\begin{figure}[h]
	\includegraphics[width=.9\textwidth]{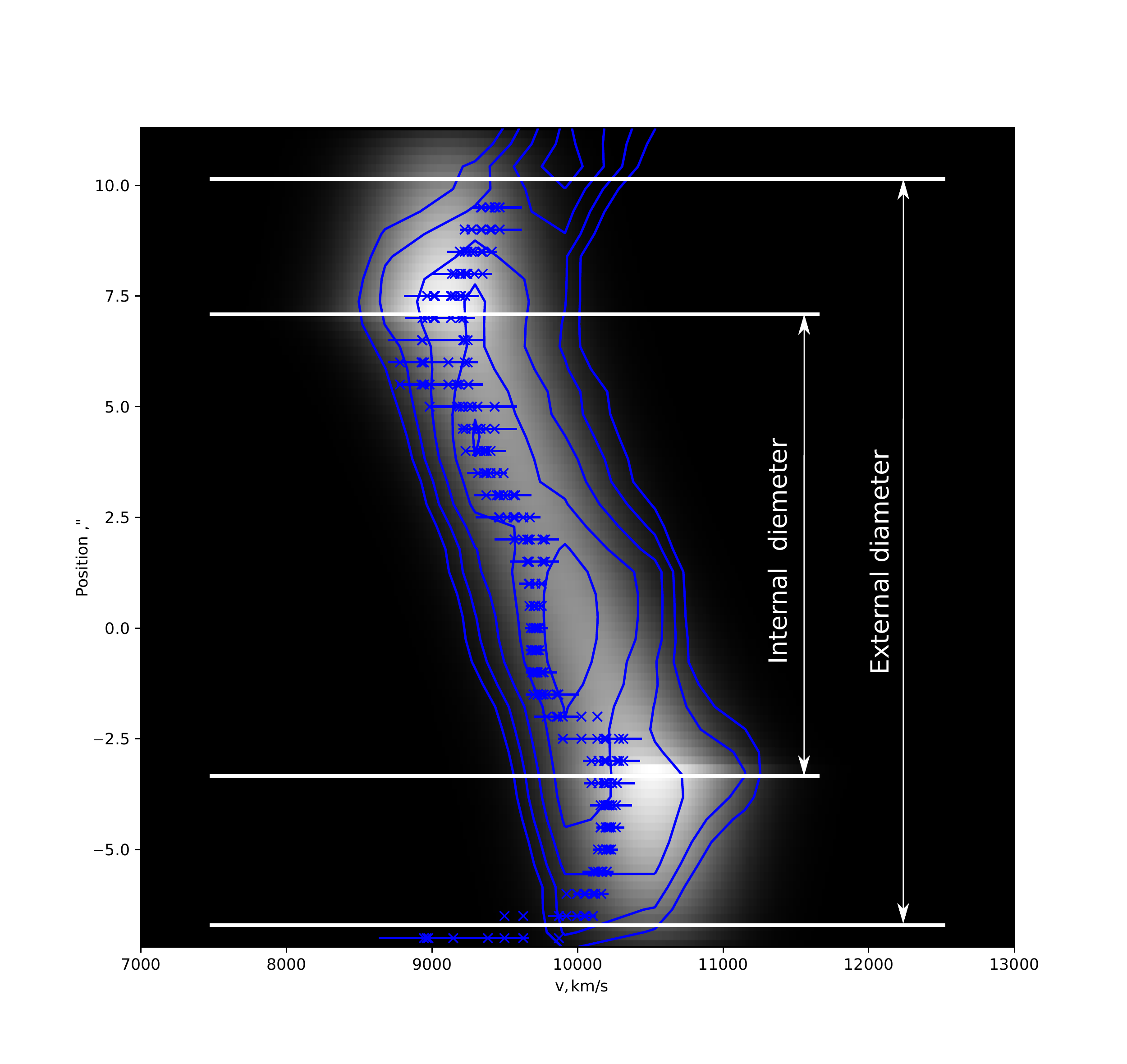}
	\caption{Position–velocity diagram in the NH$_3$ line. A model of a Keplerian torus with R$_{out}$=0.06 pc, R$_{in}$=0.04 pc, M$_{c}$=8.5~M$_\odot$ is shown in the background. The contours show the main component of the hyperfine structure of the ammonia line at levels of [2.8, 3.2, 3.7, 4.3, 5.0] K, and the x’s the velocities derived using the kNN method in the same line, with the uncertainties indicated by the horizontal segments.}
	\label{fig:nh3_pv}
\end{figure}

\begin{figure}[t]
	\includegraphics[width=.9\textwidth]{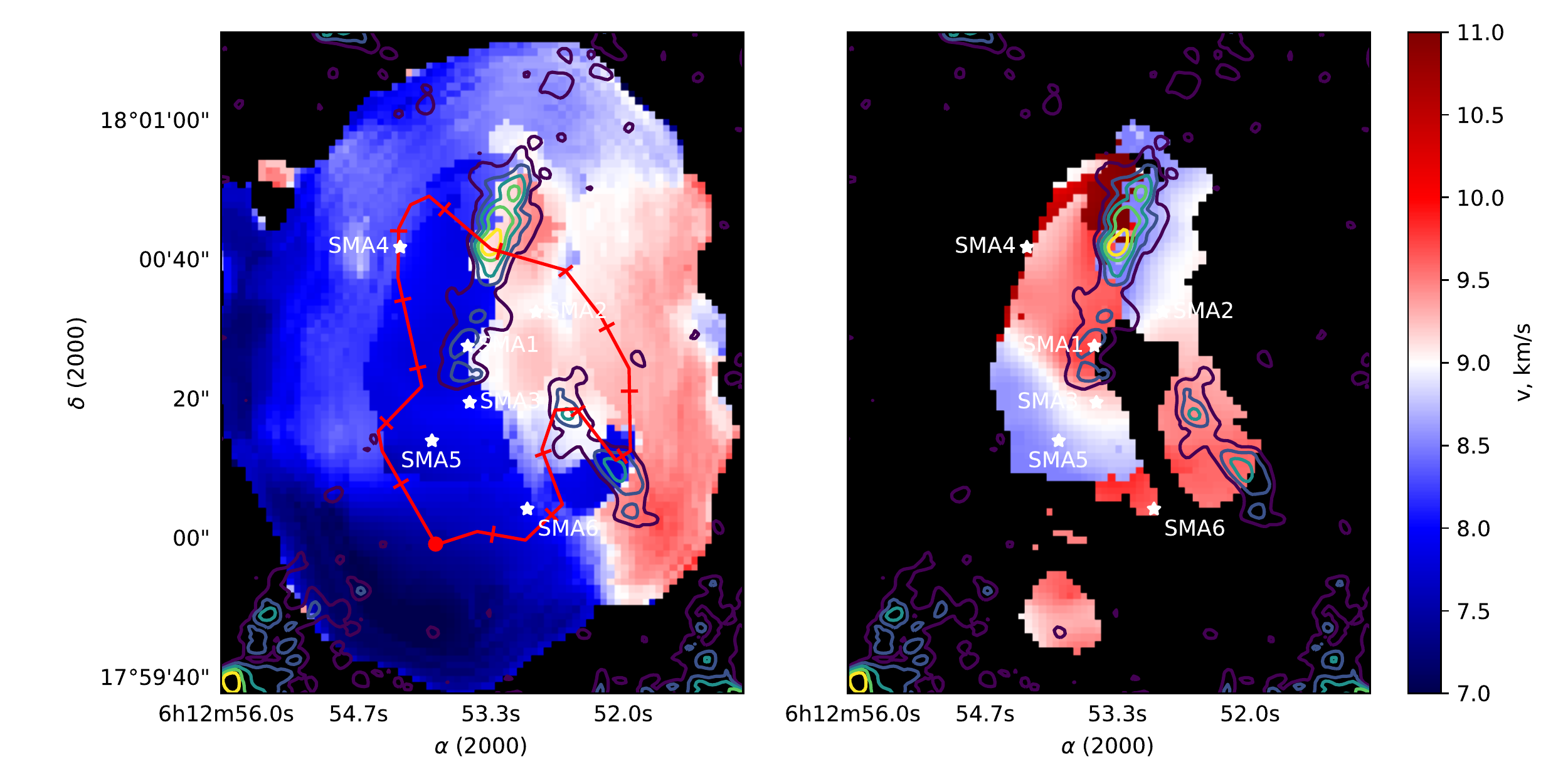}
	\caption{Maps of the Doppler velocity of the spectral components of the \nh line derived from the combined SMA and IRAM data. The left and right images show the velocities of the blue and red components in the spectra. The superposed contours show the NH$_3$ intensity at levels of [4., 5.3, 6.7, 8.1, 9.5] K. The white stars show the positions of clumps according to the data of \citep{zinhr}. The red curve shows the path of the spectrogram section in Fig. \ref{fig:pvcore}, beginning at the circle, marked every 10$''$}
	\label{fig:kn2h}
\end{figure}

\begin{figure}[h]
	\includegraphics[width=.7\textwidth]{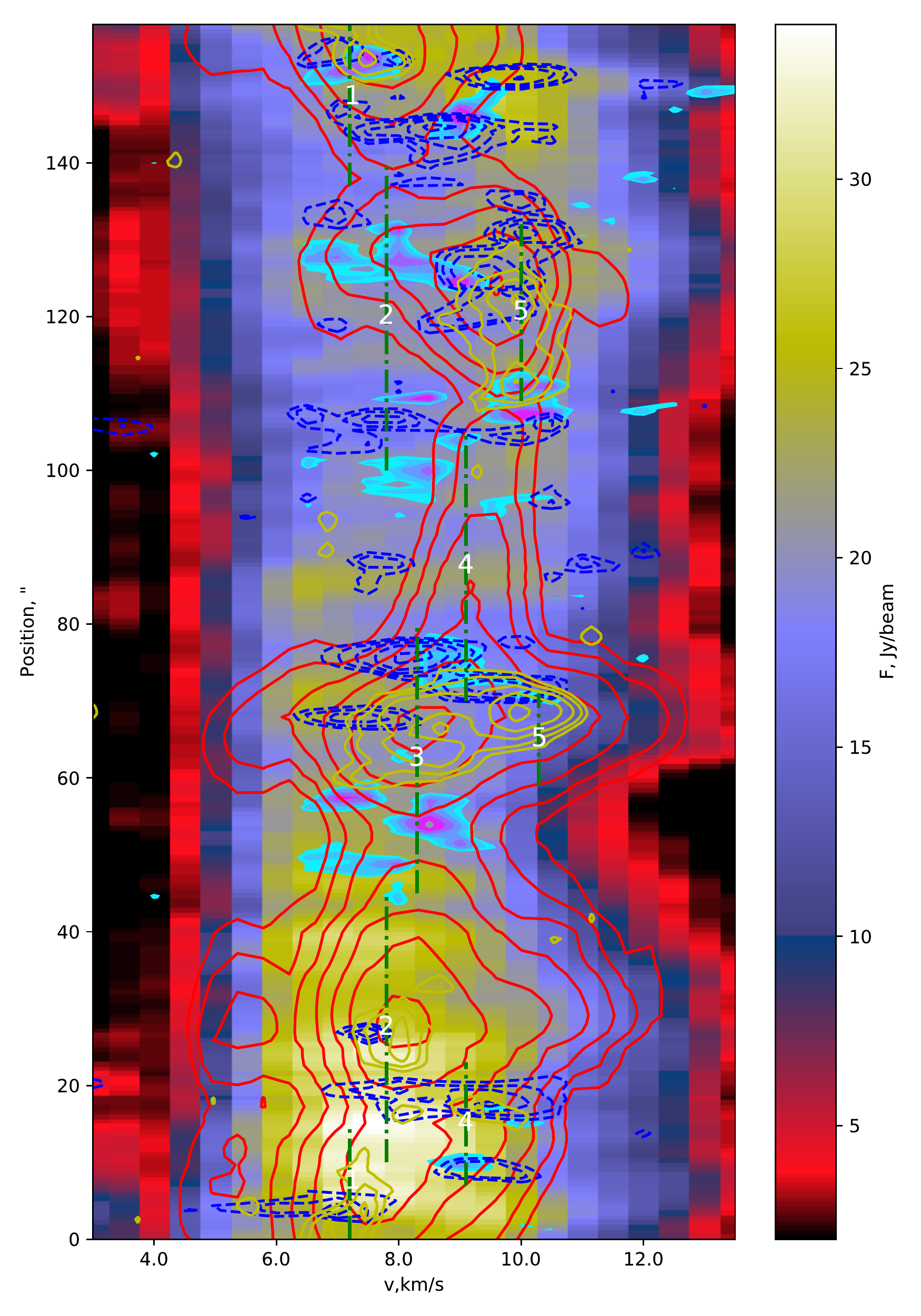}
	\caption{Spectral slice along the path shown in Fig. \ref{fig:kn2h}. The background shows the combined IRAM + SMA intensity	for the CO(2---1) line, the red contours the \nh intensity at levels of [6, 9, 12, 16, 23, 31, 43, 58] Jy/beam, the blue dashed contours the emission in the \so line at levels of [0.18, 0.2, 0.35, 0.5, 0.7, 1] Jy/beam, and the filled contours in shades of blue–violet the intensity in the \co line at levels of [0.17, 0.23, 0.3, 0.38, 0.49] Jy/beam. The vertical dot--dashed lines and numbers indicate the positions of kinematic components identified in Fig. \ref{fig:nvh}.}
	\label{fig:pvcore}
\end{figure}

\begin{figure}[h]
	\includegraphics[width=.9\textwidth]{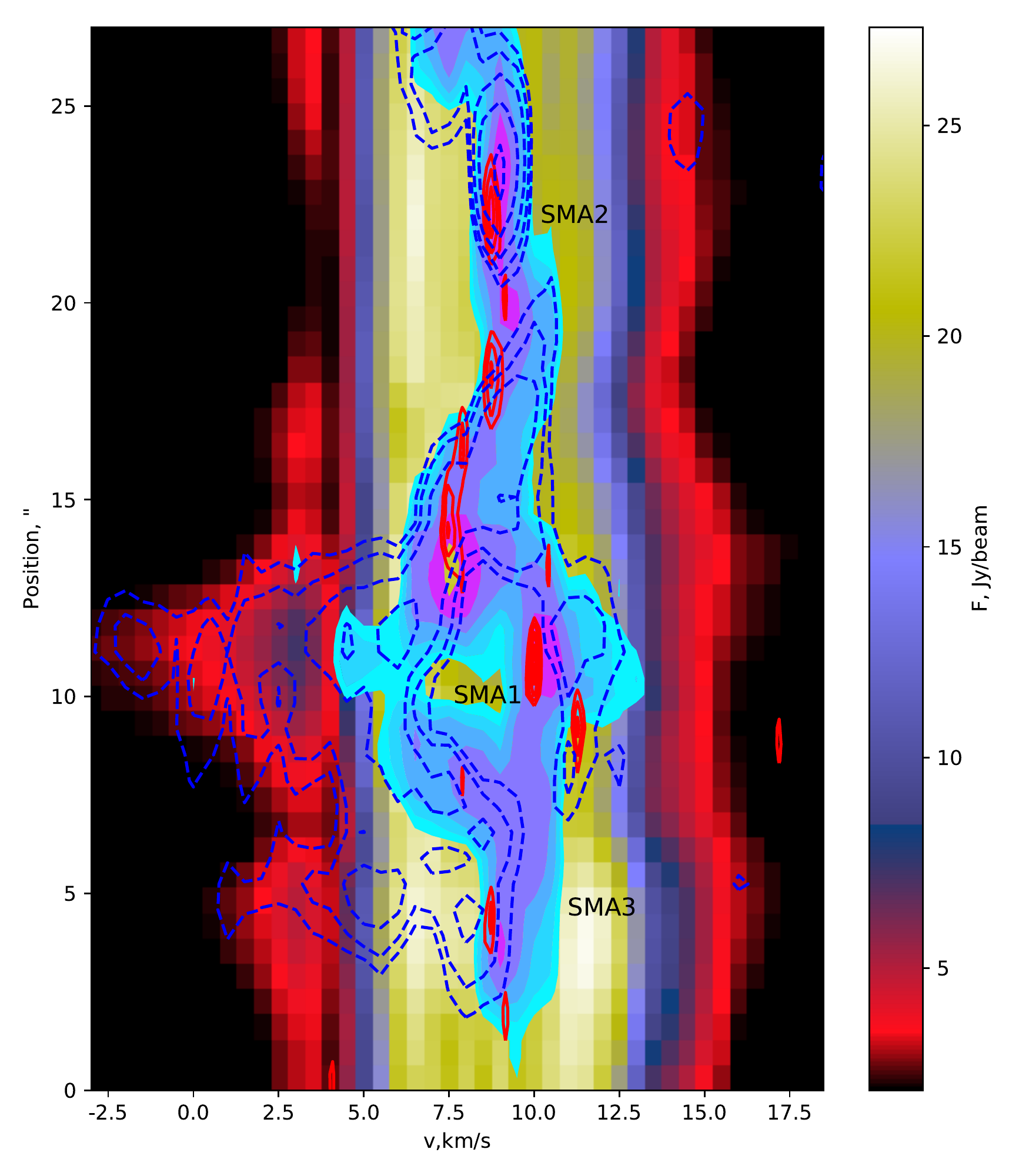}
	\caption{Spectral slice along the clumps SMA3-1-2. The background shows the CO(2---1) intensity, the filled contours in shades of blue–violet the C 18 O(2–1) emission at levels of [0.2, 0.3, 0.4, 0.6, 0.8] Jy/beam, the dashed blue contours the \so emission at levels of [0.1, 0.14, 0.18, 0.2, 0.28, 0.35, 0.44, 0.55] Jy/beam, and the red contours the DCO$^+$ (4---3) emission at levels of [0.37, 0.41, 0.45, 0.49, 0.53, 0.57, 0.62, 0.68] Jy/beam.}
	\label{fig:psma123}
\end{figure}

\begin{figure}[h]
	\includegraphics[width=0.9\textwidth]{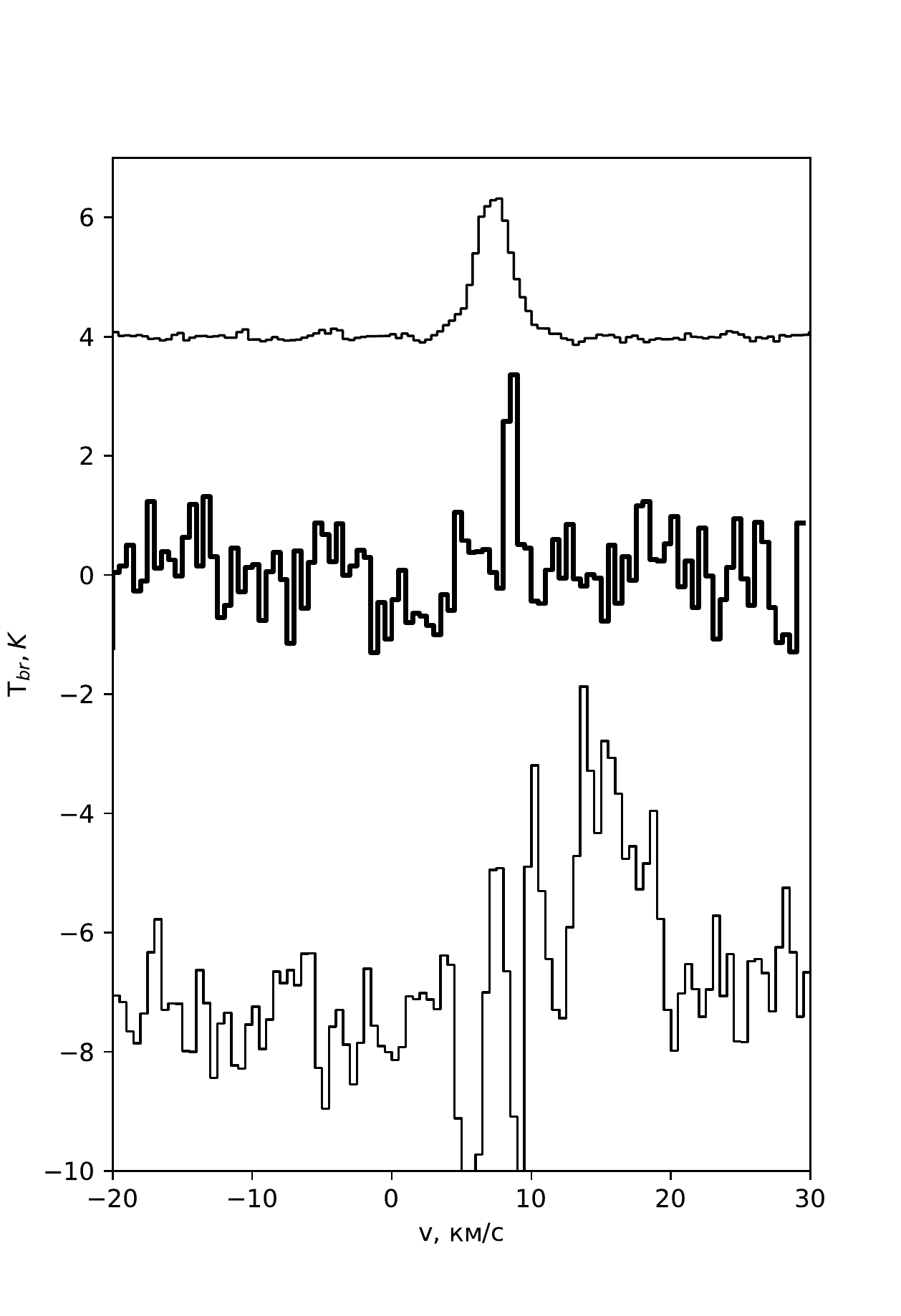}
	\caption{\nh (IRAM + SMA), DCO$^+$ (4---3), and CO(2---1) spectra in the direction of the peak DCO$^+$ emission, (-7$''$, 34$''$) relative to the phase center.}
	\label{fig:dcospecs}
\end{figure}

\begin{figure}[t]
	\includegraphics[width=0.9\textwidth]{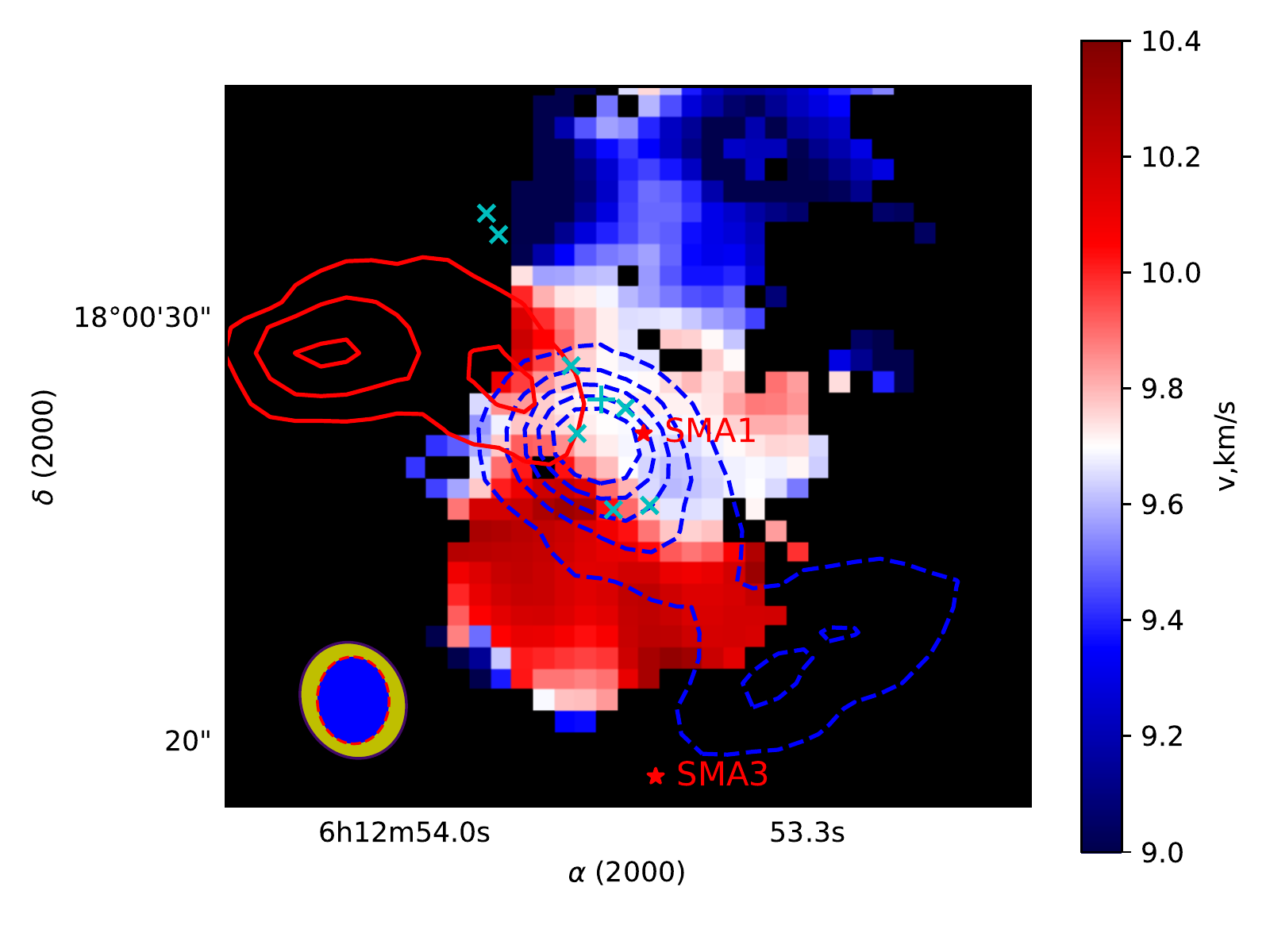}
	\caption{Map of the Doppler velocity from the fit to the \nhh line (color scale). The contours show the integrated intensity in the CO(2---1) wings, associated with the bipolar outflow, at levels of [26.5, 35.5] (solid contours) and [-18.5, -8] (dashed contours). The sizes of the VLA (outer ellipse) and SMA (inner ellipse) antenna beams are shown in the lower left-hand corner.
		The star shows the position of the clump SMA1, and the $\times$ and + symbols the positions of maser sources according to \citep{kurtz}
		(Class I methanol) and \citep{cyg2007} (H$_2$O).}
	\label{fig:nh3}
\end{figure}

The position–velocity diagram in the (1, 1) ammonia line is presented in Fig.~\ref{fig:nh3_pv}. Such a velocity profile is characteristic for a Keplerian torus (see, e.g., \citep{torus,t2}) with its plane of rotation facing us, with inner and outer radii R$_{in}\approx$ 8000 AU and R$_{out}\approx$ 12000 AU, and a central mass of M$_{c}~\approx$~8.5 M$_\odot/sin^2(i)$, where is the angle between the torus axis and the direction toward the observer. The shape of the ammonia emission contours and the linear radial dependence of the velocity in the inner part of the torus testify that the inclination of the torus is close to 90$^\circ$. We constructed a position–velocity diagram similarly to the method proposed in the appendix of \citep{pvd}, but without taking into account the radial velocity. We compared the shape of the observed and model contours in the position–velocity diagram. The sizes of the outer and inner radii can also be traced in the observed position–velocity diagram, as a linear and Keplerian part of the diagram (Fig.~\ref{fig:nh3_pv}). The inhomogeneous radial distribution of the ammonia intensity, and also the presence of emission at distances from the clump exceeding the outer radius of the torus, are noteworthy. Ammonia emission shifted by 0.8 km/s is observed to the north (the positive direction in Fig. 13).

Figure~\ref{fig:gridmap}a shows that the \co emission between the clumps SMA1 and SMA3 is continuous spatially and also in frequency. In addition, we found that SMA1 consists of two individual clumps, which we will refer to as SMA1-NE and SMA1-SW. SMA1-NE is brighter in \co and has a Doppler velocity of 7 km/s, while SMA1-SW has a Doppler velocity of 9.8 km/s. According to the \co data, their coordinates are $\alpha$(2000)=06$^h$12$^m$53.715$^s\pm$0.005$^s$ $\delta$(2000)=18$^\circ$00$'$27.73$''\pm$0.07$''$ abd $\alpha$(2000)=06$^h$12$^m$53.588$^s\pm$0.007$^s$ $\delta$(2000)=18$^\circ$00$'$26.2$''\pm$0.1$''$, respectively.

\subsection{Physical properties}

\subsubsection{Analysis of methanol observations}
\label{ch3oh}

Emission in seven methanol lines was detected in S255N: three Class I maser lines, one Class II maser line, and three quasi-thermal lines (see Table~\ref{tab1}). However, to all appearances, the emission of the Class II maser line is thermal.

We detected a bright source with the coordinates 06$^h$12$^m$53.70$^s$ +18$^\circ$00$'$24.7$'$ and with a velocity of 11.2 km/s in the $8_{-1}$--$7_{0}$ and $9_{-1}$--$8_{0}$ methanol transitions. The spectra of this source are shown in Fig.~\ref{fig:maser}. The intensity of the lines is $\approx$7 Jy/beam, appreciably higher than the intensity in the direction of the outflow with coordinates 06$^h$12$^m$54.2$^s$ +18$^\circ$00$'$14.9$'$. This large difference in flux densities suggests that this is a maser source. Maser emission at the frequencies for these transitions has also been reported, for example, in \citep{m1,m2,m3}. The maser source is localized at the southeastern boundary of the clump SMA1 and a bipolar outflow, in both space and frequency, as can be seen in Fig. \ref{fig:pvcore}. We detected only one bright source, rather than the two reported in \citep{kurtz}. Class I emission is also observed for the other maser sources reported in \citep{kurtz}. However, the insufficient sensitivity hinders drawing unambiguous conclusions about the nature of the molecular excitation. The line is fairly narrow and shows signs of maser emission in the directions toward points 5, 6, and 7, but the shape of the spectrum in the direction of the blue part of the bipolar outflow resembles the observed profiles for the SO(6$_5$---5$_4$) and SiO(5---4) lines.

The positions of the points where our estimates were carried out are shown in Fig.~\ref{fig:mcco}. In most cases, the maximum methanol line emission is shifted relative to the center of the clumps presented in \citep{zinhr}. Methanol emission peak arranged along the line of the high-velocity outflow can also be distinguished \citep{wang}. The coordinates of these maxima are presented in Table~\ref{pos}.

At all peaks line profiles differ from Gaussian. Two spectral peaks at $\sim$10 km/s and $\sim$6 km/s are clearly visible at positions (1) and (2), while there is no obvious division into spectral components at positions (3) and (4). The spectra at positions (1) and (2) were fitted with two Gaussians, and estimates were obtained separately for each of the components. Only one component was fitted at positions (3) and (4), at velocities of 7 and 8 km/s, respectively. Our estimates of the physical parameters of the gas are presented in Table~\ref{params}. 

\subsubsection{Other lines}
The angular separation of the two emission peaks in the CH$_3$CN (12$_{0}$---11$_{0}$) and \co lines within SMA1, identified with NE and SW, is 1.5$''$, which corresponds to a linear size in the plane of the sky of $\sim$2700 AU at a distance of 1.78 kpc\citep{burns}.

Our re-reduction of the data for the (12---11) transitions \citep{wang,zinhr} leads to a separation of SMA1--NE and SMA1--SW in space and frequency, as can be seen in Fig.~\ref{fig:ch3cn_pv}. Rotational diagrams \citep{ch3cn} were used to estimate the kinetic temperatures of the gas (T$_k$) for SMA1--NE and SMA1--SW, which were found to be 150$\pm$5K and 25$^{+5K}_{-10K}$, respectively. We used the first five transitions of the K ladder for the estimate for SMA1--NE, and the first three transitions for SMA1--SW. The clump temperatures indicated by the analysis of the methanol data range from 20K to 100K and from 45K to 100K, respectively, close to the estimates based on the CH$_3$CN data. The coordinates of the IR source indicated in \citep{spitzer} are appreciably closer to SMA1--NE than to SMA1--SW.

We constructed a map of the distribution of T$_k$ based on the (2, 2) and (1, 1) ammonia lines (Fig.~\ref{fig:tkin}). Typical values lie in the range 10-60 K. Figure~\ref{fig:gridmap} shows that the ammonia emission is associated with the SW clump. Temperature estimates for SMA1--SW based on the NH$_3$ data are close to the estimates from the CH$_3$CN data, and there is no indication of hot gas associated with SMA1--NE from the ammonia data. The DCN(3---2) and DCO$^+$ (4---3) emission to the south of the clumps also correlates with the region of low temperatures around SMA1---SW, but there is no DCN emission to the north. The DCO+ emission toward the north (Fig.~\ref{fig:gridmap}) is associated with gas interacting with SMA1--NE, since it lies in space and frequency inside the part of the region of \co emission where there is no ammonia. The temperature of the gas in the direction of SMA2 indicated by the ammonia data grows to 54K. A cool region with a temperature of about 10K with its peak at the coordinates 6$^h$12$^m$53$^s$, 18$^\circ$01$'$01$''$ is also observed at the northern boundary of the map. It is striking that the temperature grows to the East and West of the ammonia emission peak in S255N--NH3, consistent with the ratio of the line intensities for these two transitions at these coordinates. We used the velocity distribution of the temperatures indicated by the ammonia data to estimate the temperature range characteristic for each fragment, presented in Table \ref{tbl:vcmp}. The estimates of the kinetic temperatures based on the methanol and ammonia data are similar at points 5, 8, and NH$_3$ of Table~\ref{pos}.

The estimated mass of the filament joining the clumps SMA2, SMA1, and SMA3 is $\sim$8M$_\odot$. We used the data for the \co line to calculate this mass. The column density of C$^{18}$O was obtained using equation (15.28) of \citep{Rohlfs2004}, and the mass of gas was calculated from the abundance of this isotope of CO, 1.7$\times$10$^{-7}$ \citep{Freeking1982}.

\section{Discussion}
The core S255N has a fairly complex kinematic structure. To all appearances, it is being compressed by expanding HII regions\citep{bieging}. The region of \nh emission, which traces dense gas, is located closely between the radiation-heated dust observed in the IR around the ionized regions S255 and S257. The large-scale kinematic fragments are oriented parallel to the boundary of the gas and dust envelopes, but there is no clear indication of the nature of these fragments. According to \citep{spitzer}, 13 Class I and II protostellar IR sources were detected in the core. Seven large protostellar clumps with masses of 1--16 M$_\odot$ were also detected in the core, which are partially identified with IR sources. The presence of so many sources provides evidence of active formation of both low-mass and high-mass stars in the core. The absence of velocity profiles characteristic of compression suggests that the core has not yet made the transition to the periphery-collapse phase.

Note that the different spatial scales traced by different molecules have a mutal kinematic structure. However, several characteristic features are observed. The outer, dispersed regions of the core traced by the CO (2---1) line contain a component at a velocity of $\sim$8 km/s, which appears to be related to unperturbed gas surrounding the core. In addition, three directions for variations of the velocity can be distinguished from the center toward the northeast, the southwest, and the southeast. These last two correspond to the directions toward the ionized regions S255 and S257, suggesting that some kinematic fragments may come about as a consequence of an external interaction. Velocity slices for molecules tracing dense gas (NH$_3$ (1,1), \nh, \co) resemble the section in the CO(2---1) line, apart from the component at $\sim$8 km/s, and also the region around the clumps. However, not all of the gas fragments can be formed by external factors. In particular, an extended structure connecting the clumps SMA1, SMA2 and SMA3 and kinematically related to these clumps can be seen in the C$^{18}$O line. In addition, two fragments are observed along the line of sight in the direction of the blue wing of the bipolar outflow from SMA1 in the N$_2$H$^+$ and NH$_3$ lines; one of these is not associated with dispersed gas of the core, since it is not represent in the CO(2---1) line.

The N$_2$H$^+$ NH$_3$ C$^{18}$O and SO data suggest that the matter in the core is distributed fairly inhomogeneously, with kinematically distinct interacting gas components being present. Clumps are not present in all of these, as can be seen in the velocity in Fig.\ref{fig:nvh}.

\begin{figure}[h]
	\includegraphics[width=0.9\textwidth]{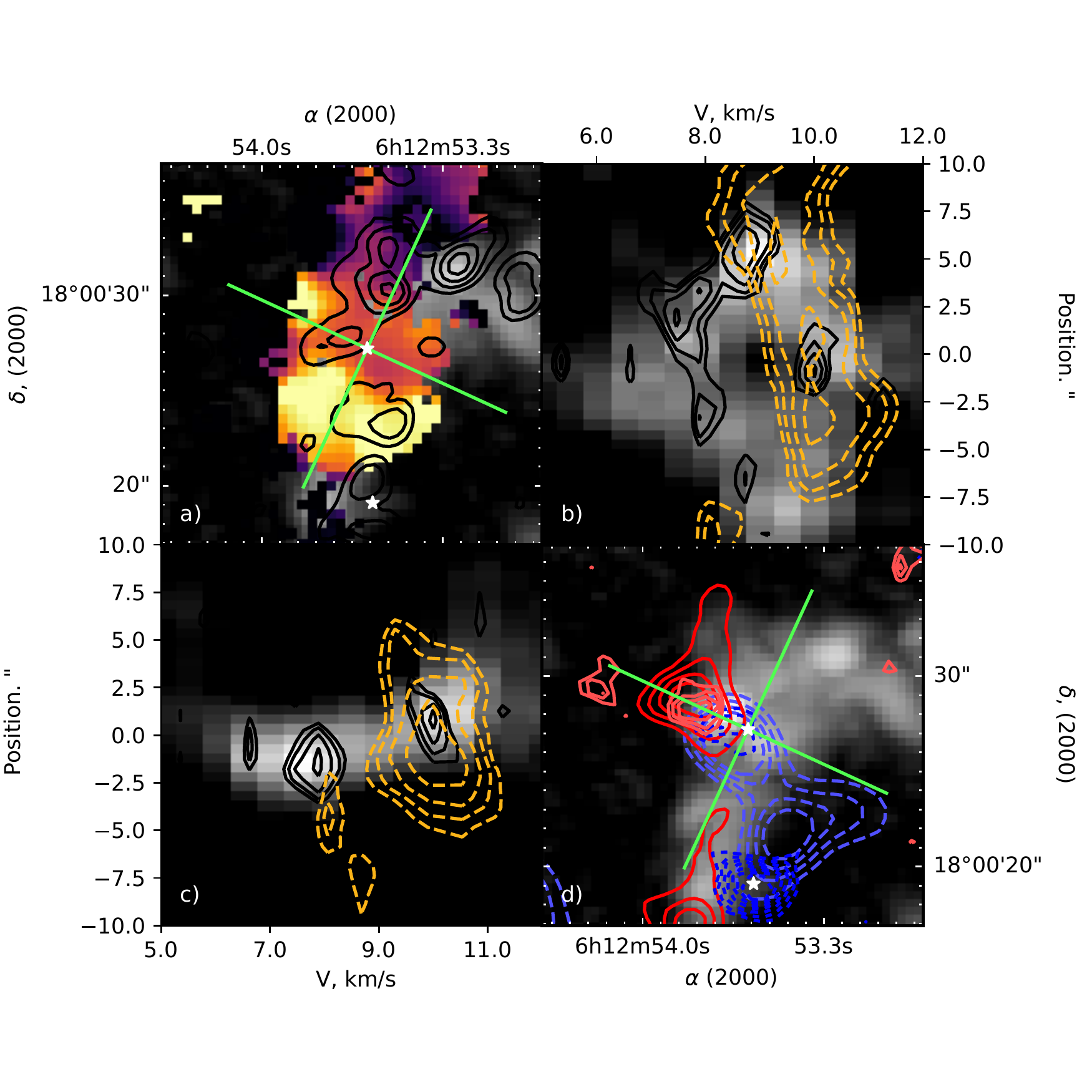}
	\caption{Spatial–kinematic structure of the clump SMA1 and its vicinity. (a) map of the \co intensity (background),
		with a map of the ammonia velocity shown in shades of yellow--violet. The black contours shows the DCO$^+$ (4---3) intensity at
		levels of [0.4, 0.55, 0.7, 0.85, 1] Jy/beam, and the green lines show the direction of the slice of the position–velocity diagram
		along and perpendicular to the direction of variations in the torus velocity (panels (b) and (c), respectively). The stars show
		the positions of clumps. (b) and (c) Position–velocity diagram along and perpendicular to the torus. The background shows
		the C$^{18}$O emission, the solid black contours the DCO$^+$ emission, and the dashed orange contours the \nhh intensity.
		(d) Map of the C$^{18}$O intensity; the solid red contours show the CO line in the red wing (pale contours are at levels of [35, 40],
		bold contours at levels of [16, 26] km/s), and the dashed blue contours the CO line in the blue wing (the large pale contours
		are at levels of [-32.5, -37.5] and the smaller contours at levels of [-5, 2] km/s}
	\label{fig:gridmap}
\end{figure}

\begin{figure}[t]
	\includegraphics[width=0.9\textwidth]{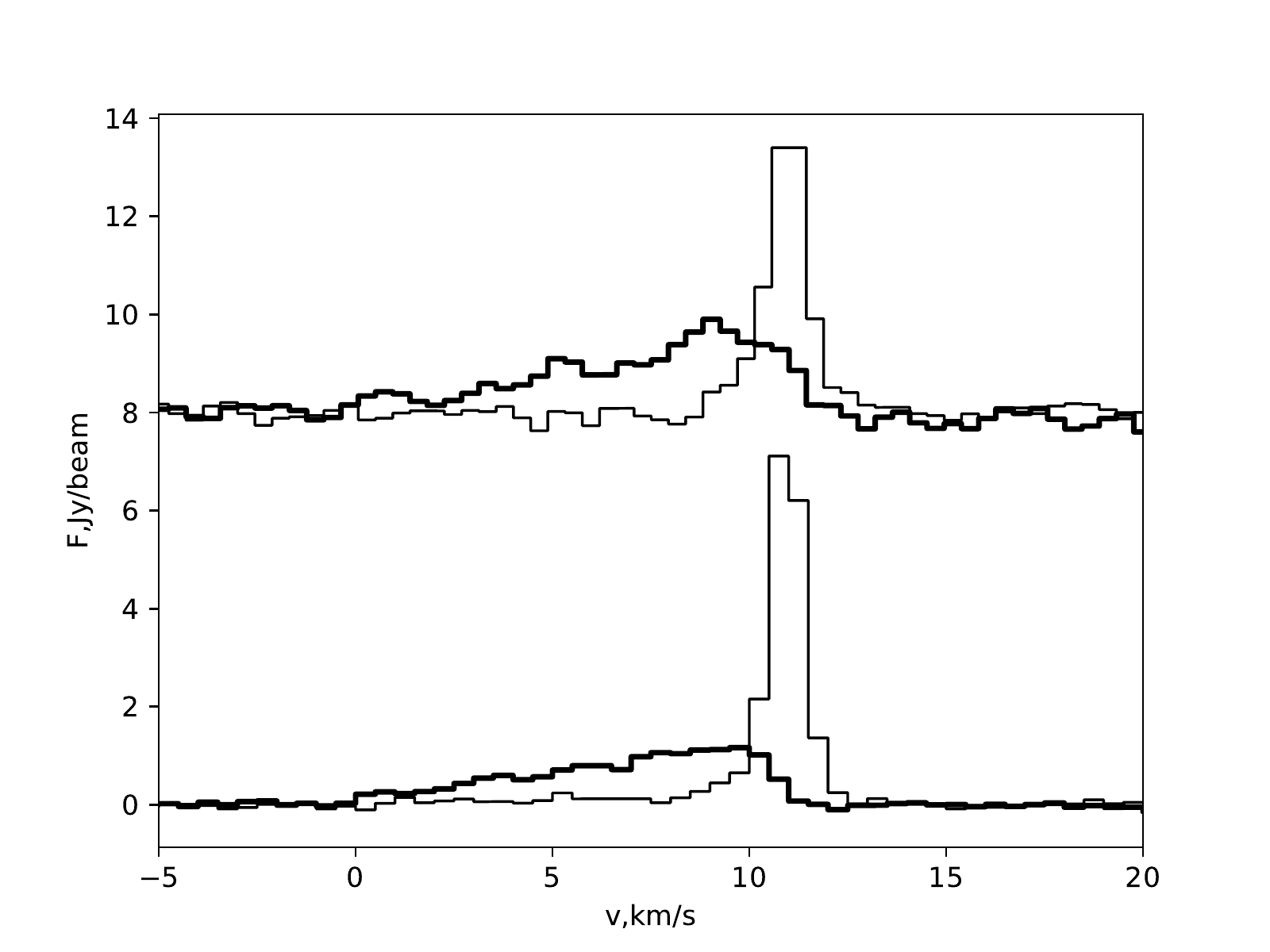}
	\caption{Methanol spectra for the transitions $8_{-1}$--$7_{0}$, $9_{-1}$--$8_{0}$ (lower, shifted by 8 Jy/beam), in the direction of the
		point with coordinates 06$^h$12$^m$53.69$^s$ +18$^\circ$00$'$25.0$'$ (J2000) (the maser spot, thin curves) and 06$^h$12$^m$54.2$^s$ +18$^\circ$00$'$14.9$'$
		(J2000) (the bipolar outflow, bold curves).}
	\label{fig:maser}
\end{figure}

\begin{figure}[t]
	\includegraphics[width=.9\textwidth]{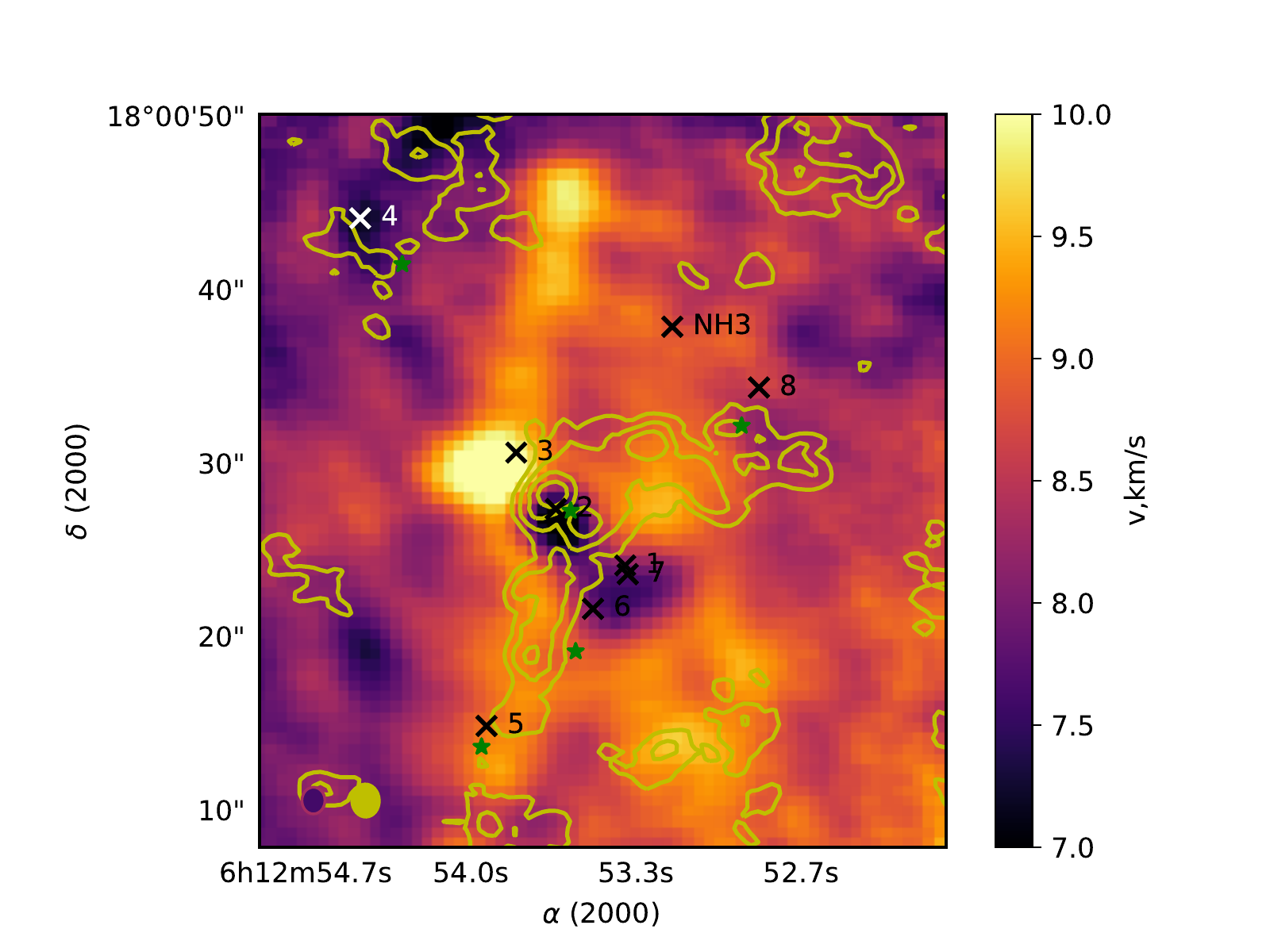}
	\caption{Map of the first moment of the CO(2---1) line. The contours show the \co intensity at levels of [0.4, 0.6, 0.8,
		1, 1.2] K. The stars show the positions of clumps according to \citep{zinhr}, and the x’s with numbers the positions of the peak methanol
		emission presented in Table ~\ref{pos}.}
	\label{fig:mcco}
\end{figure}

\begin{figure}[ht]
	\includegraphics[width=.9\textwidth]{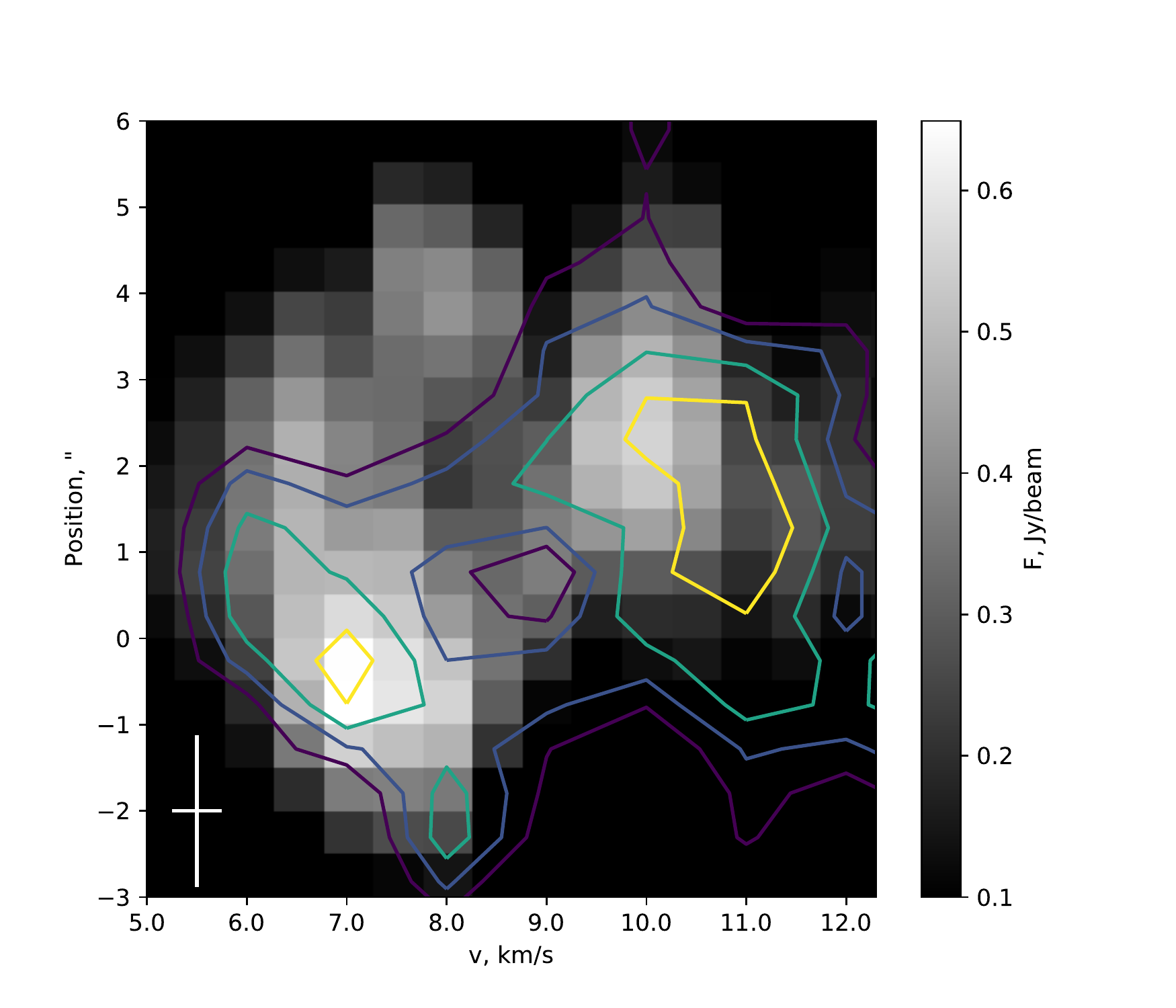}
	\caption{CH$_3$CN  (12$_0$-11$_0$) position-velocity diagram along the direction SMA1--NE~$\longrightarrow$~SMA1--SW (contours). The
		background shows the diagram for the \co line. The contours are drawn for levels of [0.07, 0.09, 0.12, 0.15, 0.2] Jy/beam. The synthesized beam of the SMA at 220.74 GHz and the width of the spectral channels are shown in the lower left corner.}
	\label{fig:ch3cn_pv}
\end{figure}

\begin{figure}[h]
	\includegraphics[width=0.9\textwidth]{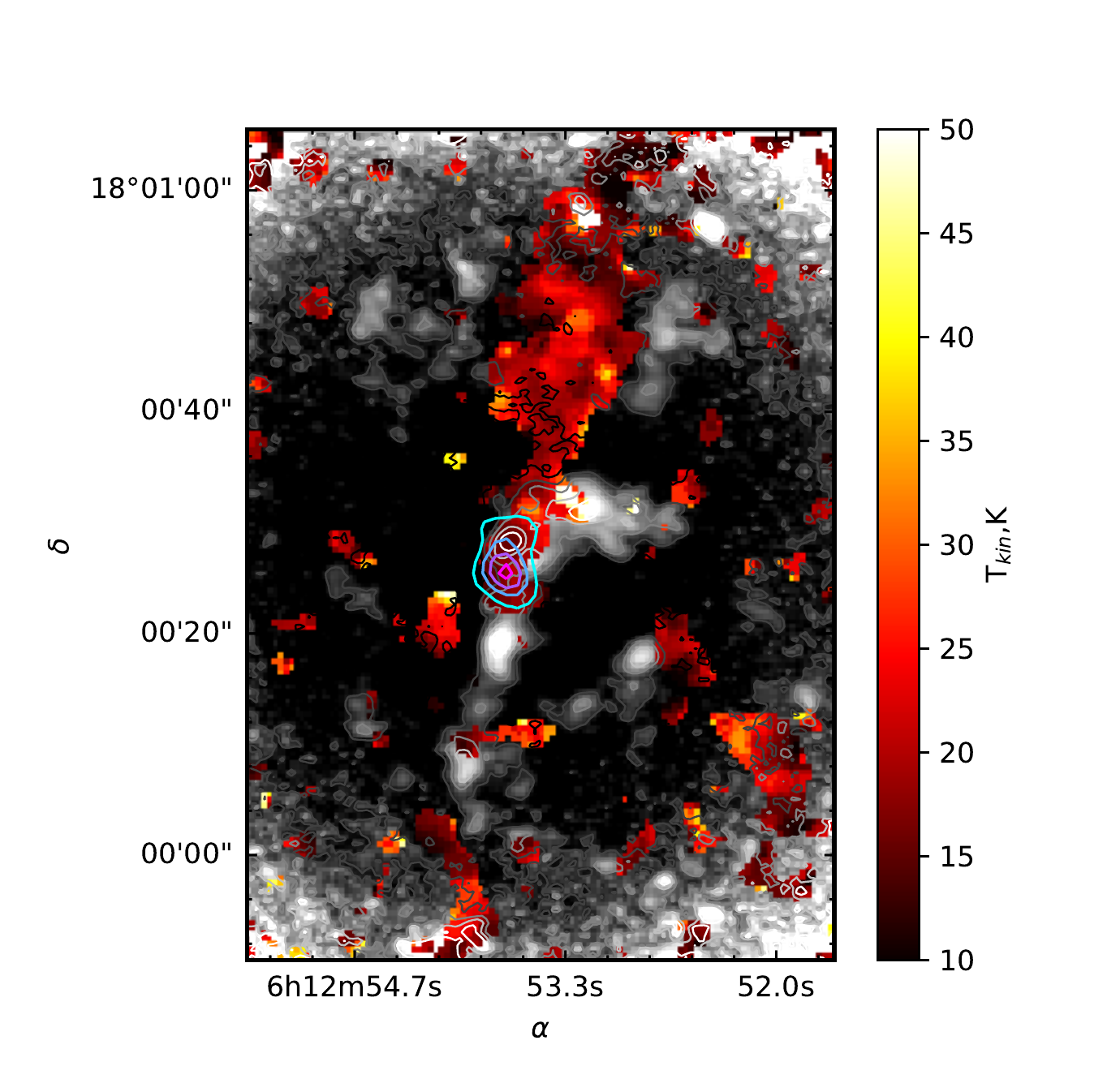}
	\caption{Map of the distribution of the kinetic temperature from the (1, 1) and (2, 2) transitions for ammonia. The shades of
		orange show the temperature estimates, the background and gray scale contours the \co intensity at levels of [0.2, 0.4, 0.6,
		0.8, 1] Jy/beam, and the blue contours the integrated DCN intensity ([-3 : 19] km/s). The contours correspond to levels of [0.7, 1.5, 2.3, 3] Jy/beam m/s.}
	\label{fig:tkin}
\end{figure}

The dense, compact gas observed in \co has an elongated structure. This emission is appreciably correlated with the emission in the \nh line, but the portion that is not is located in the region of overlap of the kinematic fragments of the core. Figure~\ref{fig:gridmap} shows that SMA1 is surrounded by matter mainly to the north and south, as can be seen in the position–velocity diagram. The line spectrum at the currently available resolution varies smoothly, separating into two components toward SMA1 (NE and SW) -- a sign of interaction of the elongated structure with the clumps. It is also important that the emission at 8-9 km/s is appreciably weaker at the center of SMA than in the filament. This may indicate that a large fraction of the filament gas in the vicinity of the clumps is interacting with this filament (Fig.~\ref{fig:pvcore}).

\begin{figure}[h]
	\includegraphics[width=0.9\textwidth]{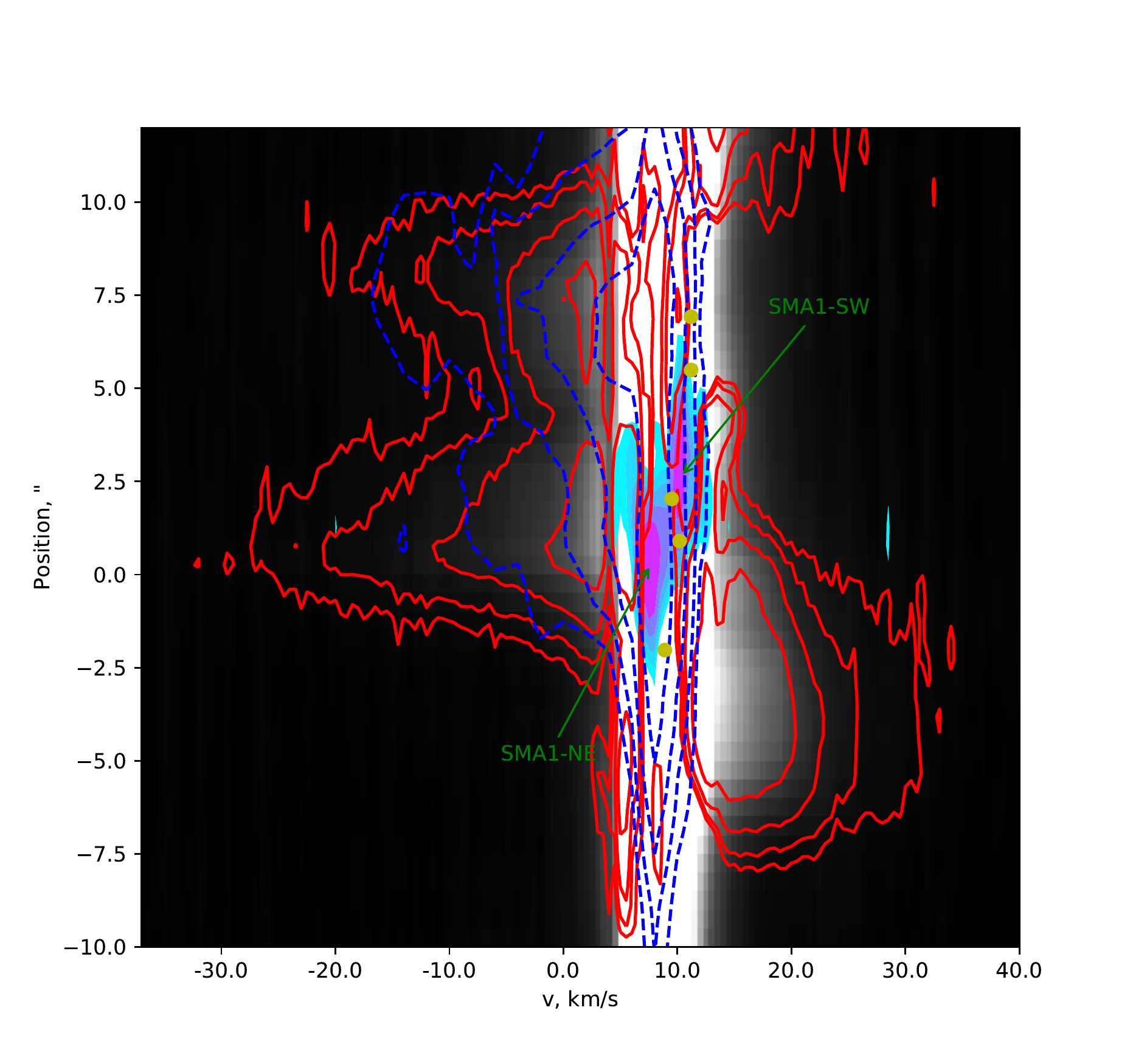}
	\caption{Position–velocity diagram along the bipolar outflow, denoted by the blue line in Fig. \ref{fig:gridmap}(a). The background shows the
		combined data in the CO (2---1) line, the red contours the interferometric data in this same line, the dashed blue contours the
		combined data in the SiO (5---4) line, and the filled contours in shades of blue–violet the data for the \co line. The filled yellow
		circles show the positions of methanol masers according to \citep{kurtz} and Section \ref{ch3oh}, and the filled yellow square the position of
		the water maser presented in \citep{cyg2007}.}
	\label{fig:outflow}
\end{figure}

The C$^{18}$O, CH$_3$CN, CH$_3$OH and DCO$^+$ data unambiguously indicate the presence of two clumps, which were not resolved earlier and have been interpreted as the object S255N SMA1. We have also continued to use the notation SMA1 when referring to the most evolved, kinematically rich central region of the core. It is noteworthy that the mean velocity of the detected giant torus coincides with the velocity of SMA1--SW, although its size is much larger than the distance between NE and SW (in the plane of the sky). Compact regions of DCO$^+$ emission that are kinematically related to both clumps are observed. An appreciably fraction of the DCO$^+$ emission is located outside the clumps, but inside the torus, correlating with C$^{18}$O (Fig. \ref{fig:gridmap}). The NE clump is associated with a bipolar outflow (Fig.~\ref{fig:outflow}) observed in the wings of CO(2---1), SO(6$_{5}$---5$_{4}$) and SiO(5---4) lines, and also with a water maser \citep{cyg2007} and methanol masers from \citep{kurtz} and detected by us. Regions of DCO$^+$ emission are present in the gas interacting with NE. This effect is not observed for SW. It is noteworthy that SW is located practically on the line of the bipolar outflow from NE. This feature could quite plausibly result due to the action of the outflow on the gas in the core. However, the relevance of the torus in this picture remains unclear. Signs of the torus are also observed in the \co data. Asymmetry of the CO velocity profiles relative to NE can be traced in Fig. \ref{fig:outflow}.

Analysis of the position–velocity diagram in the C$^{18}$O, SO, NH$_3$ and DCO$^+$ lines suggests that the torus traced in the ammonia lines is related to the accretion of the more extended, dense material of the elongated, filament-like structure onto the two young protostellar clumps. The position--velocity diagram in the C$^{18}$O line in the vicinity of SMA1 indicates an active transfer of matter between the clumps and the ambient gas in the rotational plane of the torus, with quasi-Keplerian radial profiles (Figs. \ref{fig:gridmap} a,b,c).

The detected torus is one of several dozen known disk-like objects around massive protostars \citep{disk2, diskoutflow, disk3,quanz,chini,sako}. However, a large fraction of known disks have smaller sizes or are located around less massive protostars. Tori that are larger than the disks and have appreciable thicknesses are also known \citep{t2,torus5,torus2,torus3,torus4,torus}. The object G28.20-0.05 is closest in terms of its size and physical parameters. The object we have discovered is large ($\sim$24 000 AU in diameter, as opposed to $\le$12 000 AU for G28.20-0.05). The two regions both host an ultracompact HII region. The temperatures for the two objects estimated from ammonia lines are similar\citep{torus2}. However, warmer, chemically rich torus-like rotating structures are also observed \citep{torus3,torus4}. The presence of a dust component in a similar object is also known \citep{quanz}. No signs of such a component have been detected in the case of S255N SMA1 \citep{ir,wang,zinhr}.

Theoretical models predict the appearance of torus-like objects that are gravitationally unstable \citep{transtorus}, together with an important role for the late accretion of matter onto a protostar as a transitional structure \citep{diskf}. The detection of a torus is consistent with the hypothesis of the formation of massive stars via disk accretion \citep{accretion,massive}. Observations of the region with better angular resolution and simultaneously higher sensitivity are needed for estimates of the accretion rates and detailed studies of the interaction of the torus and the ambient gas.

\section{Conclusion}
We have considered the kinematic structure of the S255N core on various spatial scales, from the outer layers of the core to dense clumps in the core. Our analysis has led to the following discoveries.
\begin{enumerate}
	\item The core does not show characteristic signs of either isotropic collapse or rotation. Overall, the large-scale motions are in agreement with those described in \citep{lev}, however, the structure of the core is very inhomogeneous.
	\item At least five extended parts of the core are observed, within which the gas velocities are similar.
	\item Large fragments of the core ($\sim$1/3 the size of the core itself) are oriented parallel to the boundaries of the HII regions S255 and S257.
	
	\item A large fraction of smaller fragments is associated with the SMA1 clump. One of these fragments is located to the southeast of SMA1 along the line of a bipolar outflow, and is elongated along this outflow. Another fragment connecting the clumps SMA3-2-1 is also elongated, and it is fairly massive ($\sim$~8M$_\odot$). 
	\item We have detected a circumstellar torus rotating around SMA1, with inner radius R$_{in}\approx$8000 AU and outer radius R$_{out}\approx$ 12000 AU. The rotation profile is characteristic of Keplerian motion around a central mass of $\sim$ 8.5/$sin^2(i)$ M$_\odot$.
	\item SMA1 is spatially resolved into two separated clumps, NE and SW. They actively interact with the torus and a filament. One of the clumps is fairly cool ($\sim$25 K), while the other is much hotter ($\sim$150 K). The bipolar outflow is associated with the hot source.
	\item A Class I maser source was detected in the CH$_3$OH $4_{2}$--$3_{1}$, $8_{-1}$--$7_{0}$ and $9_{-1}$--$8_{0}$ lines, with coordinates  $\alpha$(2000)=06$^h$12$^m$53.69$^s$ $\delta$(2000)=+18$^\circ$00$'$25.0$'$ and radial velocity $\approx$11.2 km/s. This source is located at the northeast boundary of the bipolar outflow from SMA1--NE.
	\item The kinetic temperature and number density of the gas in the clumps and outflows vary in the ranges 15--220 K and 10$^4$ -- 10$^5$ cm$^{-3}$ . The filling factor for the sources in the outflows is low, suggesting strong fragmentation of the gas.
	\item The temperature of the gas in the core is distributed inhomogeneously both in terms of the parameters of different fragments and inside the fragments themselves. Fragments that do not contain clumps have lower temperatures and smaller temperature ranges.
	\item We detected a clump with the coordinates (7$''$, -34$''$ ) relative to SMA1, together with a related high-velocity outflow.
\end{enumerate}  

\section{Acknowledgments}

We thank L.E. Pirogov and A.V. Lapinov for useful discussions of material and methods. We also thank the referee for thoughtful comments that have helped improve this paper. 

This work was supported by the Russian Foundation for Basic Research (grants 16-32-00873, 16-02-00761, 15-02-06098), the Russian Science Foundation (grant 17-12-01256), decree No. 211 of the Government of the Russian Federation (contract 02.A03.21.0006), and the Ministry of Education and Science of the Russian Federation (grant RK AAAA-A17-117030310283-7).

petez@ipfran.ru


\begin{thebibliography}{48}
\providecommand{\natexlab}[1]{#1}
\providecommand{\url}[1]{\texttt{#1}}
\expandafter\ifx\csname urlstyle\endcsname\relax
  \providecommand{\doi}[1]{doi: #1}\else
  \providecommand{\doi}{doi: \begingroup \urlstyle{rm}\Url}\fi

\bibitem[{Mapelli}(2017)]{rotcluster}
M.~{Mapelli}.
\newblock {Rotation in young massive star clusters}.
\newblock \emph{\mnras}, 467:\penalty0 3255--3267, May 2017.
\newblock \doi{10.1093/mnras/stx304}.

\bibitem[{Sokolov} et~al.(2017){Sokolov}, {Wang}, {Pineda}, {Caselli},
  {Henshaw}, {Tan}, {Fontani}, {Jim{\'e}nez-Serra}, and {Lim}]{obs}
V.~{Sokolov}, K.~{Wang}, J.~E. {Pineda}, P.~{Caselli}, J.~D. {Henshaw}, J.~C.
  {Tan}, F.~{Fontani}, I.~{Jim{\'e}nez-Serra}, and W.~{Lim}.
\newblock {Temperature structure and kinematics of the IRDC G035.39-00.33}.
\newblock \emph{\aap}, 606:\penalty0 A133, October 2017.
\newblock \doi{10.1051/0004-6361/201630350}.

\bibitem[{Richardson} et~al.(1985){Richardson}, {White}, {Gee}, {Griffin},
  {Cunningham}, {Ade}, and {Avery}]{oldie}
K.~J. {Richardson}, G.~J. {White}, G.~{Gee}, M.~J. {Griffin}, C.~T.
  {Cunningham}, P.~A.~R. {Ade}, and L.~W. {Avery}.
\newblock {Submillimetre line and continuum observations of the S255 molecular
  cloud}.
\newblock \emph{\mnras}, 216:\penalty0 713--733, October 1985.
\newblock \doi{10.1093/mnras/216.3.713}.

\bibitem[{Pirogov} et~al.(2003){Pirogov}, {Zinchenko}, {Caselli}, {Johansson},
  and {Myers}]{lev}
L.~{Pirogov}, I.~{Zinchenko}, P.~{Caselli}, L.~E.~B. {Johansson}, and P.~C.
  {Myers}.
\newblock {N$_{2}$H$^{+}$(1-0) survey of massive molecular cloud cores}.
\newblock \emph{\aap}, 405:\penalty0 639--654, July 2003.
\newblock \doi{10.1051/0004-6361:20030659}.

\bibitem[{Bieging} et~al.(2007){Bieging}, {Peters}, {Vilaro}, {Schlottman}, and
  {Kulesa}]{bieging}
J.~H. {Bieging}, W.~L. {Peters}, B.~V. {Vilaro}, K.~{Schlottman}, and
  C.~{Kulesa}.
\newblock {Sequential star formation in the Sh 254-258 molecular cloud: HHT
  maps of CO J=3-2 and 2-1 emission}.
\newblock In B.~G. {Elmegreen} and J.~{Palous}, editors, \emph{Triggered Star
  Formation in a Turbulent ISM}, volume 237 of \emph{IAU Symposium}, page 396,
  2007.
\newblock \doi{10.1017/S1743921307001846}.

\bibitem[{Wang} et~al.(2011){Wang}, {Beuther}, {Bik}, {Vasyunina}, {Jiang},
  {Puga}, {Linz}, {Rod{\'o}n}, {Henning}, and {Tamura}]{wang}
Y.~{Wang}, H.~{Beuther}, A.~{Bik}, T.~{Vasyunina}, Z.~{Jiang}, E.~{Puga},
  H.~{Linz}, J.~A. {Rod{\'o}n}, T.~{Henning}, and M.~{Tamura}.
\newblock {Different evolutionary stages in the massive star-forming region
  S255 complex}.
\newblock \emph{\aap}, 527:\penalty0 A32, March 2011.
\newblock \doi{10.1051/0004-6361/201015543}.

\bibitem[{Ojha} et~al.(2011){Ojha}, {Samal}, {Pandey}, {Bhatt}, {Ghosh},
  {Sharma}, {Tamura}, {Mohan}, and {Zinchenko}]{ojha}
D.~K. {Ojha}, M.~R. {Samal}, A.~K. {Pandey}, B.~C. {Bhatt}, S.~K. {Ghosh},
  S.~{Sharma}, M.~{Tamura}, V.~{Mohan}, and I.~{Zinchenko}.
\newblock {Star Formation Activity in the Galactic H II Complex S255-S257}.
\newblock \emph{\apj}, 738:\penalty0 156, September 2011.
\newblock \doi{10.1088/0004-637X/738/2/156}.

\bibitem[{Zinchenko} et~al.(2012){Zinchenko}, {Liu}, {Su}, {Kurtz}, {Ojha},
  {Samal}, and {Ghosh}]{zinhr}
I.~{Zinchenko}, S.-Y. {Liu}, Y.-N. {Su}, S.~{Kurtz}, D.~K. {Ojha}, M.~R.
  {Samal}, and S.~K. {Ghosh}.
\newblock {A Multi-wavelength High-resolution study of the S255 Star-forming
  Region: General Structure and Kinematics}.
\newblock \emph{\apj}, 755:\penalty0 177, August 2012.
\newblock \doi{10.1088/0004-637X/755/2/177}.

\bibitem[{Kurtz} et~al.(2004){Kurtz}, {Hofner}, and {{\'A}lvarez}]{kurtz}
S.~{Kurtz}, P.~{Hofner}, and C.~V. {{\'A}lvarez}.
\newblock {A Catalog of CH$_{3}$OH 7$_{0}$-6$_{1}$ A$^{+}$ Maser Sources in
  Massive Star-forming Regions}.
\newblock \emph{\apjs}, 155:\penalty0 149--165, November 2004.
\newblock \doi{10.1086/423956}.

\bibitem[{Zinchenko} et~al.(2009){Zinchenko}, {Caselli}, and
  {Pirogov}]{zin2009}
I.~{Zinchenko}, P.~{Caselli}, and L.~{Pirogov}.
\newblock {Chemical differentiation in regions of high-mass star formation -
  II. Molecular multiline and dust continuum studies of selected objects}.
\newblock \emph{\mnras}, 395:\penalty0 2234--2247, June 2009.
\newblock \doi{10.1111/j.1365-2966.2009.14687.x}.

\bibitem[{Minier} et~al.(2007){Minier}, {Peretto}, {Longmore}, {Burton},
  {Cesaroni}, {Goddi}, {Pestalozzi}, and {Andr{\'e}}]{minier}
V.~{Minier}, N.~{Peretto}, S.~N. {Longmore}, M.~G. {Burton}, R.~{Cesaroni},
  C.~{Goddi}, M.~R. {Pestalozzi}, and P.~{Andr{\'e}}.
\newblock {Anatomy of the S255-S257 complex - triggered high-mass star
  formation}.
\newblock In B.~G. {Elmegreen} and J.~{Palous}, editors, \emph{Triggered Star
  Formation in a Turbulent ISM}, volume 237 of \emph{IAU Symposium}, pages
  160--164, 2007.
\newblock \doi{10.1017/S1743921307001391}.

\bibitem[{Kurtz} et~al.(1994){Kurtz}, {Churchwell}, and {Wood}]{kurtz94}
S.~{Kurtz}, E.~{Churchwell}, and D.~O.~S. {Wood}.
\newblock {Ultracompact H II regions. 2: New high-resolution radio images}.
\newblock \emph{\apjs}, 91:\penalty0 659--712, April 1994.
\newblock \doi{10.1086/191952}.

\bibitem[{Cyganowski} et~al.(2007){Cyganowski}, {Brogan}, and
  {Hunter}]{cyg2007}
C.~J. {Cyganowski}, C.~L. {Brogan}, and T.~R. {Hunter}.
\newblock {Evidence for a Massive Protocluster in S255N}.
\newblock \emph{\aj}, 134:\penalty0 346--358, July 2007.
\newblock \doi{10.1086/518740}.

\bibitem[{Burns} et~al.(2016){Burns}, {Handa}, {Nagayama}, {Sunada}, and
  {Omodaka}]{burns}
R.~A. {Burns}, T.~{Handa}, T.~{Nagayama}, K.~{Sunada}, and T.~{Omodaka}.
\newblock {H$_{2}$O masers in a jet-driven bow shock: episodic ejection from a
  massive young stellar object}.
\newblock \emph{\mnras}, 460:\penalty0 283--290, July 2016.
\newblock \doi{10.1093/mnras/stw958}.

\bibitem[{Sault} et~al.(1995){Sault}, {Teuben}, and {Wright}]{miriad}
R.~J. {Sault}, P.~J. {Teuben}, and M.~C.~H. {Wright}.
\newblock {A Retrospective View of MIRIAD}.
\newblock In R.~A. {Shaw}, H.~E. {Payne}, and J.~J.~E. {Hayes}, editors,
  \emph{Astronomical Data Analysis Software and Systems IV}, volume~77 of
  \emph{Astronomical Society of the Pacific Conference Series}, page 433, 1995.

\bibitem[{Schwab}(1984)]{selfcal}
F.~R. {Schwab}.
\newblock {Relaxing the isoplanatism assumption in self-calibration;
  applications to low-frequency radio interferometry}.
\newblock \emph{\aj}, 89:\penalty0 1076--1081, July 1984.
\newblock \doi{10.1086/113605}.

\bibitem[{M{\"u}ller} et~al.(2005){M{\"u}ller}, {Schl{\"o}der}, {Stutzki}, and
  {Winnewisser}]{cdms}
H.~S.~P. {M{\"u}ller}, F.~{Schl{\"o}der}, J.~{Stutzki}, and G.~{Winnewisser}.
\newblock {The Cologne Database for Molecular Spectroscopy, CDMS: a useful tool
  for astronomers and spectroscopists}.
\newblock \emph{Journal of Molecular Structure}, 742:\penalty0 215--227, May
  2005.
\newblock \doi{10.1016/j.molstruc.2005.01.027}.

\bibitem[{Knuth}(2006)]{knuth}
K.~H. {Knuth}.
\newblock {Optimal Data-Based Binning for Histograms}.
\newblock \emph{ArXiv Physics e-prints}, May 2006.

\bibitem[{Scargle} et~al.(2013){Scargle}, {Norris}, {Jackson}, and
  {Chiang}]{bb}
J.~D. {Scargle}, J.~P. {Norris}, B.~{Jackson}, and J.~{Chiang}.
\newblock {Studies in Astronomical Time Series Analysis. VI. Bayesian Block
  Representations}.
\newblock \emph{\apj}, 764:\penalty0 167, February 2013.
\newblock \doi{10.1088/0004-637X/764/2/167}.

\bibitem[{Astropy Collaboration} et~al.(2013){Astropy Collaboration},
  {Robitaille}, {Tollerud}, {Greenfield}, {Droettboom}, {Bray}, {Aldcroft},
  {Davis}, {Ginsburg}, {Price-Whelan}, {Kerzendorf}, {Conley}, {Crighton},
  {Barbary}, {Muna}, {Ferguson}, {Grollier}, {Parikh}, {Nair}, {Unther},
  {Deil}, {Woillez}, {Conseil}, {Kramer}, {Turner}, {Singer}, {Fox}, {Weaver},
  {Zabalza}, {Edwards}, {Azalee Bostroem}, {Burke}, {Casey}, {Crawford},
  {Dencheva}, {Ely}, {Jenness}, {Labrie}, {Lian Lim}, {Pierfederici},
  {Pontzen}, {Ptak}, {Refsdal}, {Servillat}, and {Streicher}]{astropy}
{Astropy Collaboration}, T.~P. {Robitaille}, E.~J. {Tollerud}, P.~{Greenfield},
  M.~{Droettboom}, E.~{Bray}, T.~{Aldcroft}, M.~{Davis}, A.~{Ginsburg}, A.~M.
  {Price-Whelan}, W.~E. {Kerzendorf}, A.~{Conley}, N.~{Crighton}, K.~{Barbary},
  D.~{Muna}, H.~{Ferguson}, F.~{Grollier}, M.~M. {Parikh}, P.~H. {Nair}, H.~M.
  {Unther}, C.~{Deil}, J.~{Woillez}, S.~{Conseil}, R.~{Kramer}, J.~E.~H.
  {Turner}, L.~{Singer}, R.~{Fox}, B.~A. {Weaver}, V.~{Zabalza}, Z.~I.
  {Edwards}, K.~{Azalee Bostroem}, D.~J. {Burke}, A.~R. {Casey}, S.~M.
  {Crawford}, N.~{Dencheva}, J.~{Ely}, T.~{Jenness}, K.~{Labrie}, P.~{Lian
  Lim}, F.~{Pierfederici}, A.~{Pontzen}, A.~{Ptak}, B.~{Refsdal},
  M.~{Servillat}, and O.~{Streicher}.
\newblock {Astropy: A community Python package for astronomy}.
\newblock \emph{\aap}, 558:\penalty0 A33, October 2013.
\newblock \doi{10.1051/0004-6361/201322068}.

\bibitem[Altman(1992)]{knn}
N.~S. Altman.
\newblock An introduction to kernel and nearest-neighbor nonparametric
  regression.
\newblock \emph{The American Statistician}, 46\penalty0 (3):\penalty0 175--185,
  1992.
\newblock \doi{10.1080/00031305.1992.10475879}.
\newblock URL
  \url{http://www.tandfonline.com/doi/abs/10.1080/00031305.1992.10475879}.

\bibitem[{Salii}(2006)]{sali2004}
S.~V. {Salii}.
\newblock {A database for estimating the physical parameters of molecular
  clouds by the intensities of the radio lines of methanol}.
\newblock \emph{Star Formation in the Galaxy and Beyond}, pages 146--151, 2006.
\newblock URL \url{ttp://hdl.handle.net/10995/46181}.

\bibitem[{Cragg} et~al.(2005){Cragg}, {Sobolev}, and {Godfrey}]{sobolev}
D.~M. {Cragg}, A.~M. {Sobolev}, and P.~D. {Godfrey}.
\newblock {Models of class II methanol masers based on improved molecular
  data}.
\newblock \emph{\mnras}, 360:\penalty0 533--545, June 2005.
\newblock \doi{10.1111/j.1365-2966.2005.09077.x}.

\bibitem[{Miralles} et~al.(1997){Miralles}, {Salas}, {Cruz-Gonz{\'a}lez}, and
  {Kurtz}]{ir}
M.~P. {Miralles}, L.~{Salas}, I.~{Cruz-Gonz{\'a}lez}, and S.~{Kurtz}.
\newblock {Discovery of Jets and HH-like Objects near the S255 IR Complex}.
\newblock \emph{\apj}, 488:\penalty0 749--759, October 1997.
\newblock \doi{10.1086/304713}.

\bibitem[{Chavarria} et~al.(2008){Chavarria}, {Allen}, {Hora}, {Brunt}, and
  {Fazio}]{spitzer}
L.~A. {Chavarria}, L.~E. {Allen}, J.~L. {Hora}, C.~M. {Brunt}, and G.~G.
  {Fazio}.
\newblock {Spitzer Observations of the Massive Star-forming Complex S254-S258:
  Structure and Evolution}.
\newblock \emph{\apj}, 682:\penalty0 445-462, July 2008.
\newblock \doi{10.1086/588810}.

\bibitem[{Higuchi} et~al.(2015){Higuchi}, {Hasegawa}, {Saigo}, {Sanhueza}, and
  {Chibueze}]{torus}
A.~E. {Higuchi}, T.~{Hasegawa}, K.~{Saigo}, P.~{Sanhueza}, and J.~O.
  {Chibueze}.
\newblock {Sgr B2(N): A Bipolar Outflow and Rotating Hot Core Revealed by
  ALMA}.
\newblock \emph{\apj}, 815:\penalty0 106, December 2015.
\newblock \doi{10.1088/0004-637X/815/2/106}.

\bibitem[{Beltr{\'a}n} et~al.(2014){Beltr{\'a}n}, {S{\'a}nchez-Monge},
  {Cesaroni}, {Kumar}, {Galli}, {Walmsley}, {Etoka}, {Furuya}, {Moscadelli},
  {Stanke}, {van der Tak}, {Vig}, {Wang}, {Zinnecker}, {Elia}, and
  {Schisano}]{t2}
M.~T. {Beltr{\'a}n}, {\'A}.~{S{\'a}nchez-Monge}, R.~{Cesaroni}, M.~S.~N.
  {Kumar}, D.~{Galli}, C.~M. {Walmsley}, S.~{Etoka}, R.~S. {Furuya},
  L.~{Moscadelli}, T.~{Stanke}, F.~F.~S. {van der Tak}, S.~{Vig}, K.-S. {Wang},
  H.~{Zinnecker}, D.~{Elia}, and E.~{Schisano}.
\newblock {Filamentary structure and Keplerian rotation in the high-mass
  star-forming region G35.03+0.35 imaged with ALMA}.
\newblock \emph{\aap}, 571:\penalty0 A52, November 2014.
\newblock \doi{10.1051/0004-6361/201424031}.

\bibitem[{Ohashi} et~al.(1997){Ohashi}, {Hayashi}, {Ho}, and {Momose}]{pvd}
N.~{Ohashi}, M.~{Hayashi}, P.~T.~P. {Ho}, and M.~{Momose}.
\newblock {Interferometric Imaging of IRAS 04368+2557 in the L1527 Molecular
  Cloud Core: A Dynamically Infalling Envelope with Rotation}.
\newblock \emph{\apj}, 475:\penalty0 211--223, January 1997.
\newblock \doi{10.1086/303533}.

\bibitem[{Val'tts}(1998)]{m1}
I.~E. {Val'tts}.
\newblock {The unusual methanol maser 345.01+1.79}.
\newblock \emph{Astronomy Letters}, 24:\penalty0 788--794, November 1998.

\bibitem[{Hunter} et~al.(2014){Hunter}, {Brogan}, {Cyganowski}, and
  {Young}]{m2}
T.~R. {Hunter}, C.~L. {Brogan}, C.~J. {Cyganowski}, and K.~H. {Young}.
\newblock {Subarcsecond Imaging of the NGC 6334 I(N) Protocluster: Two Dozen
  Compact Sources and a Massive Disk Candidate}.
\newblock \emph{\apj}, 788:\penalty0 187, June 2014.
\newblock \doi{10.1088/0004-637X/788/2/187}.

\bibitem[Yanagida et~al.(2014)Yanagida, Sakai, Hirota, Sakai, Foster, Sanhueza,
  Jackson, Furuya, Aikawa, and Yamamoto]{m3}
Takahiro Yanagida, Takeshi Sakai, Tomoya Hirota, Nami Sakai, Jonathan~B.
  Foster, Patricio Sanhueza, James~M. Jackson, Kenji Furuya, Yuri Aikawa, and
  Satoshi Yamamoto.
\newblock Alma observations of the irdc clump g34.43+00.24 mm3: 278 ghz class
  i methanol masers.
\newblock \emph{The Astrophysical Journal Letters}, 794\penalty0 (1):\penalty0
  L10, 2014.
\newblock URL \url{http://stacks.iop.org/2041-8205/794/i=1/a=L10}.

\bibitem[{Marti}n Hernandez et~al.(2008){Marti}n Hernandez, {Bik}, {Puga},
  {N{\"u}rnberger}, and {Bronfman}]{ch3cn}
N.~L. {Marti}n Hernandez, A.~{Bik}, E.~{Puga}, D.~E.~A. {N{\"u}rnberger}, and
  L.~{Bronfman}.
\newblock {Spatially resolved near-infrared spectroscopy of the massive
  star-forming region IRAS 19410+2336}.
\newblock \emph{\aap}, 489:\penalty0 229--243, October 2008.
\newblock \doi{10.1051/0004-6361:200810336}.

\bibitem[{Rohlfs} and {Wilson}(2004)]{Rohlfs2004}
K.~{Rohlfs} and T.~L. {Wilson}.
\newblock \emph{{Tools of radio astronomy}}.
\newblock Springer, 2004.

\bibitem[{Frerking} et~al.(1982){Frerking}, {Langer}, and
  {Wilson}]{Freeking1982}
M.~A. {Frerking}, W.~D. {Langer}, and R.~W. {Wilson}.
\newblock {The relationship between carbon monoxide abundance and visual
  extinction in interstellar clouds}.
\newblock \emph{\apj}, 262:\penalty0 590--605, November 1982.
\newblock \doi{10.1086/160451}.

\bibitem[{Beuther} et~al.(2017){Beuther}, {Walsh}, {Johnston}, {Henning},
  {Kuiper}, {Longmore}, and {Walmsley}]{disk2}
H.~{Beuther}, A.~J. {Walsh}, K.~G. {Johnston}, T.~{Henning}, R.~{Kuiper}, S.~N.
  {Longmore}, and C.~M. {Walmsley}.
\newblock {Fragmentation and disk formation in high-mass star formation: The
  ALMA view of G351.77-0.54 at 0.06'' resolution}.
\newblock \emph{\aap}, 603:\penalty0 A10, June 2017.
\newblock \doi{10.1051/0004-6361/201630126}.

\bibitem[{Zinchenko} et~al.(2015){Zinchenko}, {Liu}, {Su}, {Salii}, {Sobolev},
  {Zemlyanukha}, {Beuther}, {Ojha}, {Samal}, and {Wang}]{diskoutflow}
I.~{Zinchenko}, S.-Y. {Liu}, Y.-N. {Su}, S.~V. {Salii}, A.~M. {Sobolev},
  P.~{Zemlyanukha}, H.~{Beuther}, D.~K. {Ojha}, M.~R. {Samal}, and Y.~{Wang}.
\newblock {The Disk-outflow System in the S255IR Area of High-mass Star
  Formation}.
\newblock \emph{\apj}, 810:\penalty0 10, September 2015.
\newblock \doi{10.1088/0004-637X/810/1/10}.

\bibitem[{Beltr{\'a}n} and {de Wit}(2016)]{disk3}
M.~T. {Beltr{\'a}n} and W.~J. {de Wit}.
\newblock {Accretion disks in luminous young stellar objects}.
\newblock \emph{\aapr}, 24:\penalty0 6, January 2016.
\newblock \doi{10.1007/s00159-015-0089-z}.

\bibitem[{Quanz} et~al.(2010){Quanz}, {Beuther}, {Steinacker}, {Linz},
  {Birkmann}, {Krause}, {Henning}, and {Zhang}]{quanz}
S.~P. {Quanz}, H.~{Beuther}, J.~{Steinacker}, H.~{Linz}, S.~M. {Birkmann},
  O.~{Krause}, T.~{Henning}, and Q.~{Zhang}.
\newblock {A Large, Massive, Rotating Disk Around an Isolated Young Stellar
  Object}.
\newblock \emph{\apj}, 717:\penalty0 693--707, July 2010.
\newblock \doi{10.1088/0004-637X/717/2/693}.

\bibitem[{Chini} et~al.(2006){Chini}, {Hoffmeister}, {Nielbock}, {Scheyda},
  {Steinacker}, {Siebenmorgen}, and {N{\"u}rnberger}]{chini}
R.~{Chini}, V.~H. {Hoffmeister}, M.~{Nielbock}, C.~M. {Scheyda},
  J.~{Steinacker}, R.~{Siebenmorgen}, and D.~{N{\"u}rnberger}.
\newblock {A Remnant Disk around a Young Massive Star}.
\newblock \emph{\apjl}, 645:\penalty0 L61--L64, July 2006.
\newblock \doi{10.1086/505862}.

\bibitem[{Sako} et~al.(2005){Sako}, {Yamashita}, {Kataza}, {Miyata}, {Okamoto},
  {Honda}, {Fujiyoshi}, {Terada}, {Kamazaki}, {Jiang}, {Hanawa}, and
  {Onaka}]{sako}
S.~{Sako}, T.~{Yamashita}, H.~{Kataza}, T.~{Miyata}, Y.~K. {Okamoto},
  M.~{Honda}, T.~{Fujiyoshi}, H.~{Terada}, T.~{Kamazaki}, Z.~{Jiang},
  T.~{Hanawa}, and T.~{Onaka}.
\newblock {No high-mass protostars in the silhouette young stellar object
  M17-SO1}.
\newblock \emph{\nat}, 434:\penalty0 995--998, April 2005.
\newblock \doi{10.1038/nature03471}.

\bibitem[{Boley} et~al.(2013){Boley}, {Linz}, {van Boekel}, {Henning}, {Feldt},
  {Kaper}, {Leinert}, {M{\"u}ller}, {Pascucci}, {Robberto}, {Stecklum},
  {Waters}, and {Zinnecker}]{torus5}
P.~A. {Boley}, H.~{Linz}, R.~{van Boekel}, T.~{Henning}, M.~{Feldt},
  L.~{Kaper}, C.~{Leinert}, A.~{M{\"u}ller}, I.~{Pascucci}, M.~{Robberto},
  B.~{Stecklum}, L.~B.~F.~M. {Waters}, and H.~{Zinnecker}.
\newblock {The VLTI/MIDI survey of massive young stellar objects . Sounding the
  inner regions around intermediate- and high-mass young stars using
  mid-infrared interferometry}.
\newblock \emph{\aap}, 558:\penalty0 A24, October 2013.
\newblock \doi{10.1051/0004-6361/201321539}.

\bibitem[{Sollins} et~al.(2005){Sollins}, {Zhang}, {Keto}, and {Ho}]{torus2}
P.~K. {Sollins}, Q.~{Zhang}, E.~{Keto}, and P.~T.~P. {Ho}.
\newblock {An Infalling Torus of Molecular Gas around the Ultracompact H II
  Region G28.20-0.05}.
\newblock \emph{\apj}, 631:\penalty0 399--410, September 2005.
\newblock \doi{10.1086/432503}.

\bibitem[{Beltr{\'a}n} et~al.(2011{\natexlab{a}}){Beltr{\'a}n}, {Cesaroni},
  {Neri}, and {Codella}]{torus3}
M.~T. {Beltr{\'a}n}, R.~{Cesaroni}, R.~{Neri}, and C.~{Codella}.
\newblock {Rotating toroids in G10.62-0.38, G19.61-0.23, and G29.96-0.02}.
\newblock \emph{\aap}, 525:\penalty0 A151, January 2011{\natexlab{a}}.
\newblock \doi{10.1051/0004-6361/201015049}.

\bibitem[{Beltr{\'a}n} et~al.(2011{\natexlab{b}}){Beltr{\'a}n}, {Cesaroni},
  {Zhang}, {Galv{\'a}n-Madrid}, {Beuther}, {Fallscheer}, {Neri}, and
  {Codella}]{torus4}
M.~T. {Beltr{\'a}n}, R.~{Cesaroni}, Q.~{Zhang}, R.~{Galv{\'a}n-Madrid},
  H.~{Beuther}, C.~{Fallscheer}, R.~{Neri}, and C.~{Codella}.
\newblock {Molecular outflows and hot molecular cores in G24.78+0.08 at
  sub-arcsecond angular resolution}.
\newblock \emph{\aap}, 532:\penalty0 A91, August 2011{\natexlab{b}}.
\newblock \doi{10.1051/0004-6361/201117200}.

\bibitem[{Cesaroni} et~al.(2007){Cesaroni}, {Galli}, {Lodato}, {Walmsley}, and
  {Zhang}]{transtorus}
R.~{Cesaroni}, D.~{Galli}, G.~{Lodato}, C.~M. {Walmsley}, and Q.~{Zhang}.
\newblock {Disks Around Young O-B (Proto)Stars: Observations and Theory}.
\newblock \emph{Protostars and Planets V}, pages 197--212, 2007.

\bibitem[{Kratter} et~al.(2010){Kratter}, {Matzner}, {Krumholz}, and
  {Klein}]{diskf}
K.~M. {Kratter}, C.~D. {Matzner}, M.~R. {Krumholz}, and R.~I. {Klein}.
\newblock {On the Role of Disks in the Formation of Stellar Systems: A
  Numerical Parameter Study of Rapid Accretion}.
\newblock \emph{\apj}, 708:\penalty0 1585--1597, January 2010.
\newblock \doi{10.1088/0004-637X/708/2/1585}.

\bibitem[{Balbus} and {Hawley}(1998)]{accretion}
S.~A. {Balbus} and J.~F. {Hawley}.
\newblock {Instability, turbulence, and enhanced transport in accretion disks}.
\newblock \emph{Reviews of Modern Physics}, 70:\penalty0 1--53, January 1998.
\newblock \doi{10.1103/RevModPhys.70.1}.

\bibitem[{Krumholz} et~al.(2009){Krumholz}, {Klein}, {McKee}, {Offner}, and
  {Cunningham}]{massive}
M.~R. {Krumholz}, R.~I. {Klein}, C.~F. {McKee}, S.~S.~R. {Offner}, and A.~J.
  {Cunningham}.
\newblock {The Formation of Massive Star Systems by Accretion}.
\newblock \emph{Science}, 323:\penalty0 754, February 2009.
\newblock \doi{10.1126/science.1165857}.

\end{thebibliography}


\newpage

\begin{table}[h]
	\caption{List of analyzed lines. The symbol $^*$ denotes data obtained in the extended and compact configurations. SMA+IRAM and SMA$^*$+IRAM denote the combined data}
	\begin{tabular}{l|c|c}
		Molecule, transition& Frequency, GHz & Telescope\\
		\hline 
		NH$_3$ (1,1) 	   		& 23.69 & VLA \\ 
		NH$_3$ (2,2)   	    & 23.72 & VLA  \\ 
		CH$_3$OH-E ($5_{1}$---$4_{2}$)& 216.94&SMA\\
		SiO (5---4)				& 217.10& SMA+IRAM \\
		CO (2---1)          		& 217.23 & SMA$^*$+IRAM\\ 
		DCN (3---2)         		& 217.23 & SMA  \\ 
		CH$_3$OH-E ($4_{2}$---$3_{1}$)       & 218.44&SMA\\
		C$^{18}$O (2---1)   		& 219.65 & SMA$^*$\\ 
		CH$_3$CN (12---11)		& 220.59 & SMA*  \\ 
		CH$_3$OH-E ($8_{0}$---$7_{1}$)       & 220.07&SMA\\
		CH$_3$OH-E ($8_{-1}$---$7_{0}$)      & 229.75&SMA\\
		CH$_3$OH-E ($3_{-2}$---$4_{-1}$)     & 230.02&SMA\\
		CH$_3$OH ($9_{-1}$---$8_{0}$)      & 278.30&SMA\\
		N$_{2}$H$^+$ (3---2) 		& 279.51& SMA+IRAM  \\ 
		DCO$^+$ (4---3)     		& 288.14 & SMA  \\ 
		SO (6$_{5}$---5$_{4}$) &288.51 & SMA$^*$\\
		C$^{34}$S (6---5)    		& 289.20& SMA  \\ 
		CH$_3$OH-E ($6_{0}$---$5_{0}$)       & 289.93&SMA\\

	\end{tabular} 
	\label{t:mol}
\end{table}

\begin{table}[]
	\centering
	\caption{Methanol lines detected at 216–290 GHz used to estimate physical conditions}
	\label{tab1}
	\begin{tabular}{@{}lcccc@{}}
		\hline
		Transition        & Frequency, GHz      & $E_{low}$,K   & $E_{up}$,K    &  Comments \\ 
		\hline
		
		$5_{1}$---$4_{2}$       & 216.945         & 37 &  47      & Class II maser \\
		$4_{2}$---$3_{1}$       & 218.440         & 27 &  37      & Class I maser \\
		$8_{0}$---$7_{1}$       & 220.078         & 78 &  88      &           \\
		$8_{-1}$---$7_{0}$      & 229.758         & 70 &  81      & Class I maser \\
		$3_{-2}$---$4_{-1}$     & 230.027         & 20 &  31      &        \\
		$9_{-1}$---$8_{0}$      & 278.304         & 88 & 102      & Class I maser \\
		$6_{0}$---$5_{0}$       & 289.939         & 40 &  53      &           \\  
		\hline
		
	\end{tabular}
\end{table}

\begin{table}[h!]
	\centering
	\caption{Coordinates of methanol emission peaks}
	\label{pos}
	\begin{tabular}{@{}lccc@{}}
		\hline
		
		N&			RA(2000)	&      	 	DEC(2000) &       	Nearest	object\\ 
		\hline
		
		1&	6:12:53.42	&	18:00:23.65	&	Near SMA1 \\
		2&	6:12:53.70	&	18:00:26.90	&	2 and 3\footnotemark[1]\\
		3&	6:12:53.86	&	18:00:30.15	&	5 and 6\footnotemark[1]\\
		4&	6:12:54.49	&	18:00:43.65	&	SMA4\\
		5&	6:12:53.98	&	18:00:14.40	&	SMA5, 7\footnotemark[1]\\
		6&	6:12:53.55	&	18:00:21.15	&	SMA3\\
		7&	6:12:53.41	&	18:00:23.15	&	SMA1\\
		8&	6:12:52.88	&	18:00:33.90	&	SMA2 \\ 
		NH$_3$&6:12:53.23  &	18:00:37.40 & NH$_3$ (1,1) peak\\ 	
		\hline

	\end{tabular}
	\footnotetext[1]{Maser spot detected at 44 GHz \citep{kurtz}}
\end{table}

\begin{table}[]
	\footnotesize
	\centering
	\caption{Physical parameters of gas from the fitting. 1$\sigma$ parameter range shown in parantheses. At position (4), the density range of 3--9 corresponds to the entire range in the database (model gives no constraints).  $ff$ is the beam filling factor. Ntr is the number of model parameters. The column "!" gives the number of lines with non-zero intensities at the indicated positions}
	\label{params}
	\begin{tabular}{@{}lcccccccc@{}}
		\hline
		N&	V$_{LSR}$(km/s)&	T$_k$,K	&lg(N$_m$/$\Delta$V)	&	lg(n)[H$_2$]		& 	lg(N$_{CH_{3}OH}$/N$_{H_2}$)	&	ff&	$\chi^2$/Ntr&	!\\ 
		\hline
		1&	9.8	&	45(40-100)	&10.50(10.00-11.00)		&	6.0(<7.5)	&   -6			&12.45& 0.90		&7\\
		1&	5.6	&	50(>25) 	&10.00( 9.50-10.50)		&	6.5(3.0-9.0)&	-8 			&1.15 &        	0.16&    	7\\
		2\footnote{SMA1--NE}&	7.6	&	30(>20) 	&11.25(10.00-12.00)		&   3.5(<7.0)	&	-6			&11   &         1.9	&     	6\\
		2\footnote{SMA1--SE}&	11.3&	60(45-100)	& 9.75( 9.50-10.00)		&	6.0(5.75-6.75)&	-6 			&77	  &        	4.4	&     	5\\		3&	7.1	&	40(30-160)	&11.25(10.5-12.5)		&	6.0(<7.5)	&	-6  		&4	  &        0.9	&     	7\\
		4&	7.8	&	120(>45) 	& 9.50(9.25-10.0)		&	6.5(3.0-9.0)&	-6       	& 69.7&        	4.4	&     	4\\
		6&	9.2&	85(50-200)	&	9.0(8.75-9.25)		& 	6.0(>5.5)	&	-6			&67.2 &      	1.32&      	4\\	
		7&	9.7&	60(40-120)	&	9.75(9.5-10.0)		&	6.0(5.5-7.0)&	-6			&65.02&      	5.09&      	6\\
		5&	7.5&	165(>50)	&8.00(<8.5)    			&6.0(5.5-7.5)	&-6      		&99.0 &	    	2.24&      	2\\
		6&	4.9&	50(10-220)	&9.25(<9.0)    			&6.0(3-9)		&-6      		&55.8 &     	0.50&      	5\\
		7&	5.4&	55(40-120)	&9.50(9.25-9.75)		&6.0(5.5-7.0)	&-6      		&65.41&      	2.22&      	6\\
		8&	4.7&	20(<30) 	&11.25(8-13)    		&3.5(3-9)		&-7      		&9.35 &     	1.05&      	3\\
		NH$_3$&9.5&	15(<25) 	&11.25(11-12)    		&5.0(3-9)		&-8				&54.8 &			1.06&	3\\	
		NH$_3$&7.6&	15(<25) 	&11.50(11.25-11.75)		&6.5(>5.5)		&-8				&77.9 &			2.60&	4\\
		\hline
	\end{tabular}
\end{table}
\begin{table}[]
	\footnotesize
	\centering
	\caption{Parameters of kinematic fragment of the core. The range of kinetic temperatures in the fragments was obtained from the (1, 1) and (2, 2) transitions of ammonia}
	\label{tbl:vcmp}
	\begin{tabular}{rccl}
		\hline
		Fragment number		& Velocity, km/s     	& T$_k$, K	&	 Comments 			\\ \hline
		1                   & $\sim$7.2					& 10-25			& SMA5 	\\
		2                   & $\sim$7.8   				& 10-40			& SMA1, SMA4	\\
		3                   & $\sim$8.3       			& 6-30,80		& SMA3            			\\
		4                   & $\sim$9.1      				& 10-53			& SMA2, SMA6	         			\\
		5       			& $\sim$10.3					& 12-23			& Oriented toward the outflow from SMA1 SMA1		\\
		\hline
	\end{tabular}
\end{table}

	

\end{document}